\documentclass[preprint,12pt,authoryear]{elsarticle}
\usepackage{fancyhdr}

\fancypagestyle{plain}{
  \fancyhf{}
  \fancyhead[R]{\thepage}

}
\usepackage{amsmath,amssymb}
\usepackage{graphicx}
\usepackage{subcaption}
\usepackage{booktabs}
\usepackage{siunitx}
\usepackage{xcolor}

\usepackage{hyperref}
\usepackage{multirow}
\usepackage{bm}
\usepackage{array}
\usepackage{float}
\usepackage{placeins}
\usepackage{longtable}
\usepackage{tabularx}
\hypersetup{colorlinks=true,linkcolor=blue,citecolor=blue,urlcolor=blue}

\journal{International Journal of Heat and Mass Transfer}

\begin{document}

\begin{frontmatter}

\title{Prescribed Wall-Heat-Flux Control of Blockage and Impulse in a Rarefied Micro-Nozzle}

\author[hsu]{Amirmehran Mahdavi}
\ead{am.me.mahdavi@gmail.com}
\ead{am.mahdavi@hsu.ac.ir}
\author[umass]{Ehsan Roohi\corref{cor1}}
\ead{roohie@umass.edu}

\address[hsu]{Department of Mechanical Engineering, Hakim Sabzevari University, Sabzevar, Iran}
\address[umass]{Department of Mechanical and Industrial Engineering, University of Massachusetts Amherst, Amherst, MA 01003, USA}
\cortext[cor1]{Corresponding author.}

\begin{abstract}
Prescribed wall heat flux provides an active route for controlling rarefied micro-nozzle flows, but its effect is governed by the coupled wall--bulk thermal response rather than by the imposed flux alone. This work uses direct simulation Monte Carlo (DSMC) simulations to study nitrogen flow in a converging--diverging micro-nozzle with cooling, adiabatic, and heating applied on the diverging wall. The imposed heat flux is scaled by the inlet kinetic-energy flux, $E=0.5\rho_i U_i^3$, giving $Q_w/E$ from $-10.5\%$ to $97.3\%$; this range spans moderate cooling, weak-to-intermediate heating, and a near-unity thermal-forcing regime. Wall and mass-flux-weighted bulk temperature profiles, film-temperature-based Nusselt and local-viscosity Brinkman-type diagnostics, gradient-length Knudsen indicators, mass-flux thickness, thrust decomposition, and proper orthogonal decomposition (POD) of signed numerical schlieren are analyzed. The results show that heating creates strong wall--bulk stratification: the wall temperature exceeds five times the inlet value, while the bulk temperature responds more gradually. Cooling cases contain locations where $T_w-T_b$ changes sign, making the local Nusselt-type response singular; the raw singular behavior is retained for diagnosis and a validity mask is used only for comparative plotting. Heating contracts the effective mass-carrying core, increasing aerodynamic blockage and reducing mass flow rate. However, strong heating increases the specific impulse from $156$ s to $201$ s because thermal and pressure-thrust augmentation outweigh the mass-flow penalty. The internal compression feature evolves into a finite viscous--thermal compression zone, and its heat-flux-parametric response remains low-dimensional, with the first two POD modes capturing more than $97\%$ of the fluctuation energy.
\end{abstract}


\end{frontmatter}

\section*{Nomenclature}
\small
\setlength{\LTpre}{0pt}
\setlength{\LTpost}{0pt}
\renewcommand{\arraystretch}{0.92}

\begin{longtable}{@{}p{0.18\textwidth}p{0.75\textwidth}@{}}
\toprule
Symbol & Meaning \\
\midrule
\endfirsthead

\toprule
Symbol & Meaning \\
\midrule
\endhead

\bottomrule
\endfoot

$A_t$ & Throat area per unit depth for the planar nozzle \\
$B_m$ & Effective aerodynamic blockage, $B_m=1-\beta_m$ \\
$Br_q^{\rm loc}$ & Local Brinkman-type imposed-flux ratio based on local VHS viscosity \\
$C_d$ & Discharge coefficient \\
$D_h$ & Local planar hydraulic height, $D_h=2h_{\rm geo}$ \\
$E$ & Reference inlet kinetic-energy flux, $E=\frac{1}{2}\rho_i U_i^3$ \\
$f$ & Molecular velocity distribution function \\
$f_{\rm dof}$ & Active molecular degrees of freedom used in the Eucken relation; $f_{\rm dof}=5$ for non-vibrating nitrogen \\
$f_{{\rm eu},N_2}$ & Nitrogen Eucken factor used for thermal conductivity, $f_{{\rm eu},N_2}=1.96$ \\
$F_{\rm mom}$ & Momentum-flux thrust contribution \\
$F_{\rm press}$ & Pressure-thrust contribution \\
$F_t$ & Total thrust \\
$g_0$ & Standard gravitational acceleration \\
$h_{\rm geo}$ & Local geometric half-height of the nozzle \\
$h_m$ & Effective mass-flux thickness \\
$I_{sp}$ & Specific impulse \\
$j_m$ & Axial mass-flux density, $j_m=\rho u$ \\
$k_B$ & Boltzmann constant \\
$k_f$ & Film-temperature-based local thermal conductivity from the VHS/Wu--Eucken relation \\
$k_{\rm ref}$ & Reference thermal conductivity, used only when a reference scaling is required \\
$Kn$ & Knudsen number \\
$Kn_{\rm GLL}$ & Gradient-length local Knudsen-number indicator \\
$L$ & Nozzle length scale \\
$L_1,L_2$ & Converging and diverging section lengths \\
$L_{\rm in}$ & Inlet half-height \\
$L_{\rm out}$ & Exit half-height \\
$L_t$ & Throat height \\
$m$ & Molecular mass of nitrogen \\
$\dot{m}$ & Mass flow rate \\
$Ma$ & Mach number \\
$N_{SQ}$ & Number of DSMC time steps used for wall heat-flux sampling before each wall-temperature update \\
$Nu_q^{\rm loc}$ & Local apparent heat-flux-based Nusselt number using temperature-dependent thermal conductivity \\
$p$ & Static pressure \\
$p_{\rm amb}$ & Ambient/back pressure used in the pressure-thrust term \\
$P_i$ & Inlet pressure \\
$P_{\rm out}$ & Outlet pressure \\
$PR$ & Pressure ratio, $PR=P_i/P_{\rm out}$ \\
$q_w$ & Sampled local wall heat flux in DSMC \\
$q_{\rm des}$ & Prescribed target wall heat flux \\
$q_w^*$ & Reduced imposed wall heat flux, when used for normalized heat-flux comparison \\
$Q_w$ & Nominal prescribed wall heat flux for a case \\
$Q_w/E$ & Dimensionless thermal forcing ratio \\
$R_F$ & Relaxation factor for wall-temperature correction \\
$s$ & Normalized coordinate along Wall-2 (Diverging wall), $s=(x-x_t)/(L-x_t)$ \\
$T$ & Gas temperature \\
$T_0$ & Inlet/reference temperature used for normalization \\
$T_b$ & Mass-flux-weighted bulk gas temperature \\
$T_f$ & Film temperature for wall-to-bulk heat-transfer scaling, $T_f=(T_w+T_b)/2$ \\
$T_{\rm ref}$ & Reference temperature in the VHS viscosity law \\
$T_w$ & Wall temperature \\
$T_{w,1}$ & Initial/isothermal wall temperature for upstream and converging walls \\
$u,v$ & Streamwise and transverse velocity components \\
$U_b$ & Mass-flux-weighted axial velocity \\
$U_i$ & Inlet velocity used in the kinetic-energy-flux scale \\
$u_{\rm slip}$ & Near-wall tangential velocity relative to the stationary wall \\
$\bm{t}_w$ & Local unit tangent vector along Wall-2 \\
$x,y$ & Physical coordinates \\
$x_t$ & Throat axial location \\
$x/L$ & Axial coordinate normalized by the nozzle length scale \\
$\beta_m$ & Normalized mass-flux thickness, $\beta_m=h_m/h_{\rm geo}$ \\
$\chi$ & Signed numerical schlieren, $\partial(\rho/\rho_0)/\partial(x/L)$ \\
$\Delta T_w$ & Wall-temperature correction in the prescribed-heat-flux feedback algorithm \\
$\Delta T_{\min}$ & Wall--bulk temperature-difference threshold used only for the Nusselt validity mask \\
$\epsilon_q$ & Regularization constant in the prescribed-heat-flux wall-temperature feedback \\
$\epsilon_Q$ & Small regularization constant in the Brinkman-type denominator \\
$\lambda$ & Molecular mean free path \\
$\mu_b$ & Bulk-temperature-based local dynamic viscosity from the VHS law \\
$\mu_f$ & Film-temperature-based local dynamic viscosity from the VHS law \\
$\mu_{\rm ref}$ & Reference dynamic viscosity used in the VHS viscosity scaling \\
$\omega$ & VHS viscosity--temperature index \\
$\rho$ & Gas density \\
$\rho_0$ & Reference density used for normalization \\
$\phi_k$ & $k$th POD spatial mode \\
$a_k$ & $k$th POD modal coefficient \\

\midrule
Acronym & Meaning \\
\midrule
CWH & Constant wall heat flux \\
DSMC & Direct simulation Monte Carlo \\
GHS & Generalized hard sphere \\
IJHMT & International Journal of Heat and Mass Transfer \\
LB & Larsen--Borgnakke internal-energy exchange model \\
MEMS & Microelectromechanical systems \\
NTC & No-time-counter collision-selection scheme \\
POD & Proper orthogonal decomposition \\
PPC & Particles per cell \\
VHS & Variable hard sphere \\
VSS & Variable soft sphere \\

\end{longtable}
\normalsize
\renewcommand{\arraystretch}{1.0}

\section{Introduction}

Micro-nozzles are enabling elements in micro-propulsion, vacuum gas handling, material processing, and microelectromechanical systems, where compact, high-speed jets are generated over characteristic dimensions comparable to molecular transport scales. In a conventional macroscopic converging--diverging nozzle, the governing picture is built around inviscid acceleration, area change, and the possible formation of shocks or shock cells under non-ideal pressure ratios. At the microscale, this picture is incomplete. The wall area per unit volume is large, viscous and thermal layers occupy a substantial fraction of the passage, gas--surface energy exchange can dominate the thermodynamic state, and the local Knudsen number can vary strongly along the nozzle. The internal flow is therefore not controlled only by pressure ratio and geometry; it is also controlled by how the wall exchanges energy and momentum with the gas.

The breakdown of continuum behavior is usually measured by the Knudsen number, $Kn=\lambda/L$, where $\lambda$ is the molecular mean free path and $L$ is a characteristic length. Slip and temperature-jump effects become important when $Kn$ approaches the slip-flow range, and transitional behavior appears when the mean free path is no longer small compared with the characteristic gradient length \citep{tsien1946,schaaf1961,karniadakis2005,sharipov2011}. In rarefied internal nozzles, this classification is complicated by strong area variation. The inlet, throat, diverging section, compression layer, and downstream buffer can each exhibit different local rarefaction levels. The resulting flow is a spatially heterogeneous kinetic-flow problem rather than a single-regime nozzle problem.

Direct simulation Monte Carlo (DSMC) remains the reference high-fidelity method for such regimes because it statistically solves the Boltzmann equation by representing molecular motion and collisions with computational particles \citep{bird1970,wagner1992}. DSMC has been applied to rarefied nozzle flows, micro-thrusters, MEMS-scale gas expansion, vacuum plumes, and microscale propulsion performance \citep{ivanov1999,wang2004,hao2005,xie2007,horisawa2008,darbandi2011,lijo2015,saadati2015,sabouri2019,mahdavi2020,sukesan2021,kosyanchuk2021,groll2023,zhang2024,sabouri2024}. These studies have shown that micro-nozzle behavior is strongly affected by viscous losses, gas--surface interaction, accommodation, and finite-thickness compression structures. They also indicate that shock-like structures in rarefied internal flows should not be treated automatically as mathematical discontinuities. A DSMC-resolved compression layer has a finite kinetic thickness, may be broadened by wall interaction, and can merge with thermal and viscous layers.

Recent rarefied-flow studies have also made clear that field-level organization is as important as integral performance. Shock-centered low-rank analysis of rarefied, specular-wall micro-nozzle data showed that internal compression structures can be compactly represented when the dominant parameter dependence is essentially a displacement and thickening of a shock-cell-like feature \citep{roohi2025shockfusion}. Rarefied bow-shock studies have shown that increasing rarefaction weakens and thickens detached shock layers, so that the compression region becomes a finite viscous--kinetic layer rather than a geometrically similar inviscid shock surface \citep{riabov1999shock_interference_rarefied,agir2022rarefaction_edney}. Together, general SciML/operator-learning frameworks and recent rarefied-flow applications indicate that reduced coordinates and training weights should be connected to identifiable physical structures; for example shocks, recirculation zones, geometric parameters, wall layers, or non-equilibrium regions; rather than introduced only as abstract numerical compression variables \citep{raissi2019,lu2021deeponet,peyvan2024,tatsios2025,roohi2026microstepdeeponet,roohi2025shockfusion}. These observations motivate the present work: if wall heat flux introduces a new coherent degree of freedom, then a reduced or design-oriented description should include wall-thermal and blockage coordinates in addition to pressure-ratio or shock-position coordinates.

The wall thermal condition introduces the central mechanism. Prescribed wall temperature has often been used as a convenient boundary condition in rarefied micro-nozzle studies, including DSMC analyses of wall-temperature effects on flow regime, slip velocity, temperature, and wall heat flux \citep{zhang2024}. However, a prescribed heat flux is more directly relevant to active thermal control, heat leakage, localized wall heating, and micro-propulsion components with finite thermal power. Heat-flux-specified gas--surface boundary treatments have been developed for DSMC by iteratively adjusting the wall temperature until the sampled molecular energy exchange matches the target wall heat flux \citep{akhlaghi2012,akhlaghi2016}. In the present work, this idea is applied to the diverging wall of a rarefied micro-nozzle and connected to wall--bulk thermal stratification, near-wall tangential slip, effective blockage, compression-zone restructuring, and propulsive performance. In a continuum solver, a heat flux is imposed through a Neumann boundary condition. In DSMC, however, the wall heat flux is a sampled molecular energy-exchange outcome. The wall temperature must be adjusted until the difference between incoming and outgoing molecular energy fluxes matches the desired value. Such a feedback boundary treatment has important physical consequences: the wall temperature is no longer a fixed input but an emergent distribution that balances the imposed heat flux with the local molecular impact statistics.

For this reason, the present paper treats prescribed wall heat flux as a coupled heat-transfer, rarefaction, and propulsion problem. Previous DSMC micro/nanochannel studies under constant wall heat flux showed that the wall--bulk temperature difference, viscous dissipation, and Brinkman scaling can produce Nusselt-number singularities and even counter-gradient heat-transfer behavior; they also emphasized that transport-property variation must be retained when the temperature field changes appreciably \citep{balaj2014cwh,balaj2015shearwork}. For molecular gases, the thermal conductivity entering a Nusselt-type diagnostic is not a single monatomic constant: translational and internal energy modes can contribute differently to heat transport. Recent kinetic-model work by Wu and co-workers for non-vibrating polyatomic gases and Rayleigh--Brillouin scattering demonstrates that Eucken factors and the decomposition of thermal conductivity are important in rarefied molecular-gas modelling \citep{wu2015polyatomic,wu2020eucken}. We therefore use diagnostics that are defensible from the available kinetic data: wall-temperature response, bulk gas-temperature response, a local VHS/Eucken Nusselt-type scaling, a local-viscosity Brinkman-type measure of the competition between viscous momentum transport and imposed heat transfer, gradient-length Knudsen-number fields, mass-flux thickness, discharge coefficient, thrust decomposition, and proper orthogonal decomposition (POD) of signed numerical schlieren. The Nusselt and Brinkman measures are used as comparative heat-transfer scalings, not as proof that continuum constitutive laws hold everywhere in the nozzle.

The central question is not simply whether heating raises temperature or whether cooling reduces it. The question is how prescribed wall heat flux changes the coupled mass, momentum, and energy conversion in a rarefied micro-nozzle. Does heating primarily improve performance by adding energy, or does it degrade performance by increasing viscous blockage? Does the internal shock-cell-like compression structure remain a distinct feature, or does it become a broad viscous--thermal compression zone? Are the resulting field changes arbitrary, or do they remain organized enough to admit a compact modal representation?

The contribution of the paper is fourfold. First, we impose controlled cooling, adiabatic, and heating conditions on the diverging wall of a rarefied nitrogen micro-nozzle using a DSMC wall-temperature feedback method. Second, we use a combined wall--bulk temperature analysis that exposes how the imposed heat flux penetrates into the flow and when the wall--bulk temperature difference changes sign. Third, we introduce field-derived aerodynamic and heat-transfer diagnostics, including $\beta_m$, $u_{\rm slip}$, $Nu_q^{\rm loc}$, $Br_q^{\rm loc}$, $Kn_{\rm GLL}$, and thrust decomposition, to connect local thermal forcing to global performance. Fourth, we use signed numerical schlieren and POD to determine whether the compression-layer response is a random DSMC scatter pattern or a coherent low-dimensional deformation. Together, these elements provide a physically stronger interpretation than a simple statement that heating moves or dissolves a shock.

The remainder of the paper is organized as follows. 
\hyperref[sec:method]{Section~\ref*{sec:method}} describes the DSMC formulation, molecular model, nozzle geometry, operating conditions, prescribed-heat-flux boundary condition, heat-flux normalization, and numerical verification. 
\hyperref[sec:configuration]{Section~\ref*{sec:configuration}} defines the field-derived diagnostics used throughout the analysis, including mass-flux thickness, effective blockage, signed numerical schlieren, gradient-length Knudsen indicators, thrust, and heat-transfer scaling measures. 
\hyperref[sec:results]{Section~\ref*{sec:results}} reports the main results, beginning with the coupled wall--bulk temperature response and the near-wall tangential slip behavior, followed by signed numerical schlieren, gradient-length Knudsen-number fields, effective blockage, mass-flux thickness, thermal and aerodynamic trade-off, Nusselt and Brinkman-type heat-transfer scalings, and POD-based modal organization. 
The final part of \hyperref[sec:results]{Section~\ref*{sec:results}} discusses the physical implications for thermal control of rarefied micro-nozzles, including the mechanisms of mass-flux-thickness contraction, impulse gain despite blockage, compression-zone restructuring, and the limitations of the present interpretation. 
\hyperref[sec:conclusions]{Section~\ref*{sec:conclusions}} summarizes the main conclusions.

\section{Numerical method and simulation setup}
\label{sec:method}

\subsection{DSMC formulation and molecular model}

The DSMC method approximates the Boltzmann equation,
\begin{equation}
\frac{\partial f}{\partial t}+\bm{c}\cdot\nabla_{\bm{x}}f=J(f,f),
\end{equation}
where $f(\bm{x},\bm{c},t)$ is the molecular velocity distribution function and $J(f,f)$ is the binary collision operator. DSMC advances the kinetic solution by separating molecular motion and intermolecular collisions over a time step smaller than the relevant collision time,
\begin{equation}
f^{n+1}\approx \mathcal{C}_{\Delta t}\mathcal{M}_{\Delta t}f^n,
\end{equation}
where $\mathcal{M}_{\Delta t}$ denotes free molecular motion and $\mathcal{C}_{\Delta t}$ denotes stochastic collision sampling within computational cells. The cell size is selected to remain smaller than the local mean free path, and the time step is selected below the local mean collision time. These requirements are most restrictive near the throat and in the internal compression region, where density, temperature, and velocity gradients are strongest.

The simulations are performed for nitrogen using the no-time-counter (NTC) collision-selection scheme, the variable-hard-sphere (VHS) molecular model, and the Larsen--Borgnakke procedure for rotational energy exchange. Gas--surface interactions at solid walls are treated using diffuse reflection with full thermal accommodation. The molecular and numerical parameters are summarized in Table~\ref{tab:dsmc_parameters}. The values reported in Table~\ref{tab:dsmc_parameters} are the final production settings selected after the grid, time-step, and particle-number sensitivity checks described in Section~\ref{sec:verification}. Macroscopic quantities are obtained by sampling statistically steady particles after the initial transient. The reported fields include density, velocity components, translational/rotational temperature, Mach number, pressure, wall heat flux, heat-transfer scalings, and Knudsen-number-related diagnostics.

\begin{table}[t]
\centering
\footnotesize
\caption{Final DSMC molecular-model and numerical parameters used in the production simulations.}
\label{tab:dsmc_parameters}
\begin{tabularx}{0.96\textwidth}{@{}lX@{}}
\toprule
Quantity & Value/model \\
\midrule
Gas & Nitrogen \\
Molecular mass, $m$ & $4.65\times10^{-26}~{\rm kg}$ \\
Reference molecular diameter, $d_{\rm ref}$ & $4.17\times10^{-10}~{\rm m}$ \\
Reference temperature, $T_{\rm ref}$ & $273~{\rm K}$ \\
Viscosity--temperature index, $\omega$ & $0.74$ \\
Molecular degrees of freedom & $5$ \\
Collision-selection scheme & No-time-counter (NTC) \\
Molecular model & Variable hard sphere (VHS) \\
Internal-energy exchange & Larsen--Borgnakke model \\
Wall reflection model & Diffuse reflection with full thermal accommodation \\
Structured computational blocks & $25\times30$ converging block, $75\times30$ diverging block, and $30\times40$ downstream buffer block \\
Collision subcells & $2\times2$ per computational cell \\
Particles per cell used in production runs & $25$ \\
Time step used in production runs, $\Delta t$ & $4\times10^{-10}~{\rm s}$ \\
Wall-temperature relaxation factor, $R_F$ & $0.05$ \\
Heat-flux regularization, $\epsilon_q$ & $10^{-8}~{\rm W\,m^{-2}}$ \\
Wall heat-flux sampling interval, $N_{SQ}$ & $1000$ time steps \\
Heat-flux convergence tolerance & $10^{-3}$ \\
Initial wall temperature, $T_{w,1}$ & $300~{\rm K}$ \\
Isothermal wall portion & Upstream and converging walls up to the throat \\
Thermally updated wall portion & Diverging-wall elements downstream of the throat \\
\bottomrule
\end{tabularx}
\end{table}

\subsection{Geometry, operating conditions, and thermal cases}

\begin{figure}[!htbp]
\centering
\includegraphics[
  width=0.82\textwidth,
  height=0.42\textheight,
  keepaspectratio,
  trim={35 85 35 75},
  clip
]{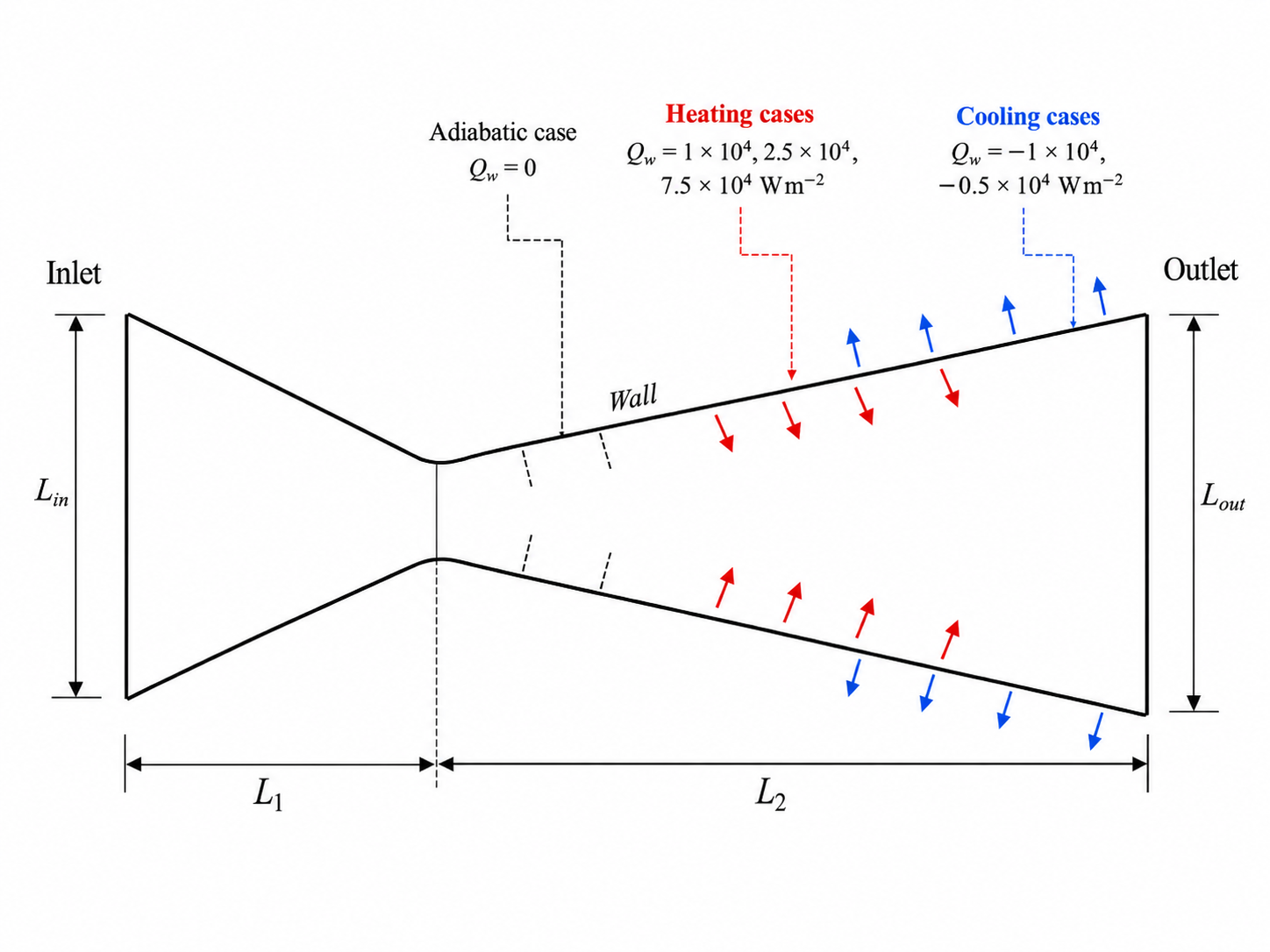}
\caption{Converging--diverging micro-nozzle and thermal boundary conditions. All prescribed heat-flux cases are applied to the diverging wall. Heating adds energy from the wall to the gas, cooling removes energy from the gas, and the adiabatic condition is represented by directionless wall-normal markers.}
\label{fig:schematic}
\end{figure}

The planar converging--diverging micro-nozzle is shown in Fig.~\ref{fig:schematic}. The inlet half-height is $L_{\rm in}=34~\mu{\rm m}$, the throat height is $L_t=15~\mu{\rm m}$, the exit half-height is $L_{\rm out}=68~\mu{\rm m}$, the converging length is $L_1=51.25~\mu{\rm m}$, and the diverging length is $L_2=153.75~\mu{\rm m}$. The inlet pressure and temperature are $P_i=1~{\rm atm}$ and $T_i=300~{\rm K}$, respectively, while the outlet pressure is $P_{\rm out}=7~{\rm kPa}$, corresponding to a pressure ratio of approximately $14.3$. The Knudsen number is defined using the inlet half-height as $Kn=\lambda/(2L_{\rm in})$. Because the geometry and boundary conditions are symmetric about the centerline, only a half-domain is simulated; full-domain visualizations are generated by mirror reflection.

The prescribed heat flux is applied only on the diverging wall, where acceleration, viscous--thermal layer growth, and the internal compression structure interact most strongly. The upstream and converging walls are kept isothermal at $T_w=300~{\rm K}$. Six thermal cases are considered: two cooling cases, $Q_w=-1.0\times10^4$ and $-0.5\times10^4~{\rm W\,m^{-2}}$; one adiabatic baseline, $Q_w=0$; and three heating cases, $Q_w=1.0\times10^4$, $2.5\times10^4$, and $7.5\times10^4~{\rm W\,m^{-2}}$. The geometric, operating, and thermal parameters are summarized in Table~\ref{tab:setup}. 

\begin{table}[t]
\centering
\footnotesize
\caption{Nozzle geometry, operating conditions, and wall-heat-flux cases.}
\label{tab:setup}
\begin{tabularx}{0.96\textwidth}{@{}lX@{}}
\toprule
Quantity & Value \\
\midrule
Inlet half-height, $L_{\rm in}$ & $34~\mu{\rm m}$ \\
Throat height, $L_t$ & $15~\mu{\rm m}$ \\
Exit half-height, $L_{\rm out}$ & $68~\mu{\rm m}$ \\
Converging length, $L_1$ & $51.25~\mu{\rm m}$ \\
Diverging length, $L_2$ & $153.75~\mu{\rm m}$ \\
Inlet pressure, $P_i$ & $1~{\rm atm}$ \\
Inlet temperature, $T_i$ & $300~{\rm K}$ \\
Outlet pressure, $P_{\rm out}$ & $7~{\rm kPa}$ \\
Pressure ratio, $P_i/P_{\rm out}$ & $14.3$ \\
Knudsen-number definition & $Kn=\lambda/(2L_{\rm in})$ \\
Thermally forced wall & Diverging wall only \\
Wall-heat-flux cases & $Q_w=\{-1.0,-0.5,0,1.0,2.5,7.5\}\times10^4~{\rm W\,m^{-2}}$ \\
\bottomrule
\end{tabularx}
\end{table}

\subsection{Prescribed heat-flux boundary condition}

In a particle method, the wall heat flux is not imposed through a continuum temperature gradient. Instead, it is the net molecular energy exchange between incident and reflected particles. Let $q_w(x)$ be the sampled wall heat flux at a wall element and $q_{\rm des}(x)$ be the prescribed target heat flux. The wall temperature is updated through \citep{akhlaghi2012,akhlaghi2016}
\begin{equation}
T_w^{n+1}(x)=T_w^n(x)+\Delta T_w(x),
\label{eq:tw_update}
\end{equation}
with a proportional relaxation correction
\begin{equation}
\frac{\Delta T_w(x)}{T_w^n(x)}
=
-R_F
\frac{q_w(x)-q_{\rm des}(x)}
{|q_{\rm des}(x)|+\epsilon_q}.
\label{eq:q_feedback}
\end{equation}

In the implementation, the wall temperature is initialized as $T_w=T_{w,1}$ for all wall elements. For the present cases $T_{w,1}=300~{\rm K}$, so the upstream and converging walls remain isothermal at the inlet temperature. When the prescribed-heat-flux option is active, the feedback correction is applied only to the diverging-wall elements downstream of the throat. Consequently, the large wall-temperature variation discussed later is not a globally imposed wall-temperature distribution; it is the converged gas--surface response required to satisfy the target heat flux on Wall-2.

Reflected molecules are sampled from a diffuse wall distribution at the local wall temperature. The most probable speed of reflected molecules is computed as $V_{\rm mp,w}=\sqrt{2k_B T_w/m}$, and the rotational energy of nitrogen molecules is resampled using the same local wall temperature through the Larsen--Borgnakke procedure. The prescribed heat-flux boundary condition, therefore, modifies both translational and rotational energy exchange at the gas--surface interface.

For convergence monitoring and post-processing, the heat-flux module records the wall collision rate, instantaneous and averaged wall heat flux, instantaneous and averaged wall temperature, wall-temperature correction, and the translational and rotational incident/reflected energy contributions. 

The relaxation factor $R_F$ damps stochastic fluctuations, while $\epsilon_q$ prevents division by zero near the adiabatic case. The heat flux is sampled over $N_{SQ}=1000$ time steps before each update so that the correction is based on statistically meaningful molecular impacts. The sign of Eq.~\eqref{eq:q_feedback} ensures that if the sampled heat transfer is smaller than the target heating value, the wall temperature increases, and if the sampled energy removal is insufficient for cooling, the wall temperature decreases. The feedback process is continued until the relative heat-flux discrepancy falls below $10^{-3}$. Thus, the final wall-temperature distribution is an outcome of the coupled gas--surface energy balance rather than a prescribed input.

\subsection{Energy scale of the imposed heat flux}

The dimensional values of the prescribed wall heat flux alone do not provide a sufficient physical justification for the thermal cases. A heat flux of $10^4~{\rm W\,m^{-2}}$ may be weak or strong depending on how much kinetic energy is advected into the micro-nozzle. Therefore, following the energy-scaling argument used in propulsion and micronozzle performance analysis \citep{rothe1971,louisos2012,groll2023}, we normalize the imposed wall heat flux by the inlet kinetic-energy flux per unit area,
\begin{equation}
E=\frac{1}{2}\rho_i U_i^3,
\qquad
\Pi_Q=\frac{Q_w}{E}.
\label{eq:q_e_def}
\end{equation}
The denominator in Eq.~\eqref{eq:q_e_def} is chosen because $\frac{1}{2}\rho_i U_i^2$ is the incoming kinetic-energy density of the gas and multiplication by $U_i$ converts it to an advective kinetic-power flux with the same units as $Q_w$ (${\rm W\,m^{-2}}$). It is evaluated from the inlet state, before any wall heat addition or removal, so it provides a fixed reference scale for all thermal cases. This choice avoids normalizing by a downstream or local quantity that is itself modified by the imposed heat flux. Thus, $\Pi_Q$ directly measures how large the imposed wall energy exchange is relative to the kinetic power carried into the nozzle.

This normalization gives a clear physical basis for the selected cooling and heating levels, and the resulting values are summarized in Table~\ref{tab:heatflux_ratio}. The two cooling cases remove approximately $5$--$10\%$ of the inlet kinetic-energy flux, so they represent moderate thermal extraction rather than an unrealistically strong refrigeration limit. The weak and intermediate heating cases add about $12\%$ and $30\%$ of the inlet kinetic-energy flux, respectively, allowing the transition from pressure-driven expansion to thermally influenced expansion to be resolved. The strongest heating case, $Q_w=7.5\times10^4~{\rm W\,m^{-2}}$, corresponds to $\Pi_Q\simeq 97.3\%$, i.e., the wall supplies thermal power comparable to the incoming kinetic-energy flux. This near-unity value is intentionally included as a limiting active-control case in which wall heating is expected to reorganize not only the temperature field but also the mass flux, compression structure, blockage, and impulse response.

\begin{table}[t]
\centering
\footnotesize
\caption{Dimensionless wall-heat-flux levels. The ratio $\Pi_Q=Q_w/E$ compares the imposed wall heat flux with the inlet kinetic-energy flux $E=\frac{1}{2}\rho_i U_i^3$ and provides the physical basis for selecting the cooling and heating cases.}
\label{tab:heatflux_ratio}
\begin{tabularx}{0.96\textwidth}{@{}lccX@{}}
\toprule
Case & $Q_w~({\rm W\,m^{-2}})$ & $\Pi_Q=Q_w/E$ & Physical interpretation \\
\midrule
Cooling 1 & $-1.0\times10^4$ & $-10.5\%$ & Moderate energy extraction \\
Cooling 2 & $-0.5\times10^4$ & $-5.06\%$ & Weak energy extraction \\
Adiabatic & $0$ & $0$ & No imposed wall energy exchange \\
Heating 1 & $1.0\times10^4$ & $11.6\%$ & Weak heat addition \\
Heating 2 & $2.5\times10^4$ & $30.2\%$ & Intermediate heat addition \\
Heating 3 & $7.5\times10^4$ & $97.3\%$ & Near-unity thermal forcing \\
\bottomrule
\end{tabularx}
\end{table}

\subsection{Numerical resolution and verification}
\label{sec:verification}

\begin{figure}[!htbp]
\centering
\includegraphics[width=1.1\textwidth,height=0.62\textheight,keepaspectratio]{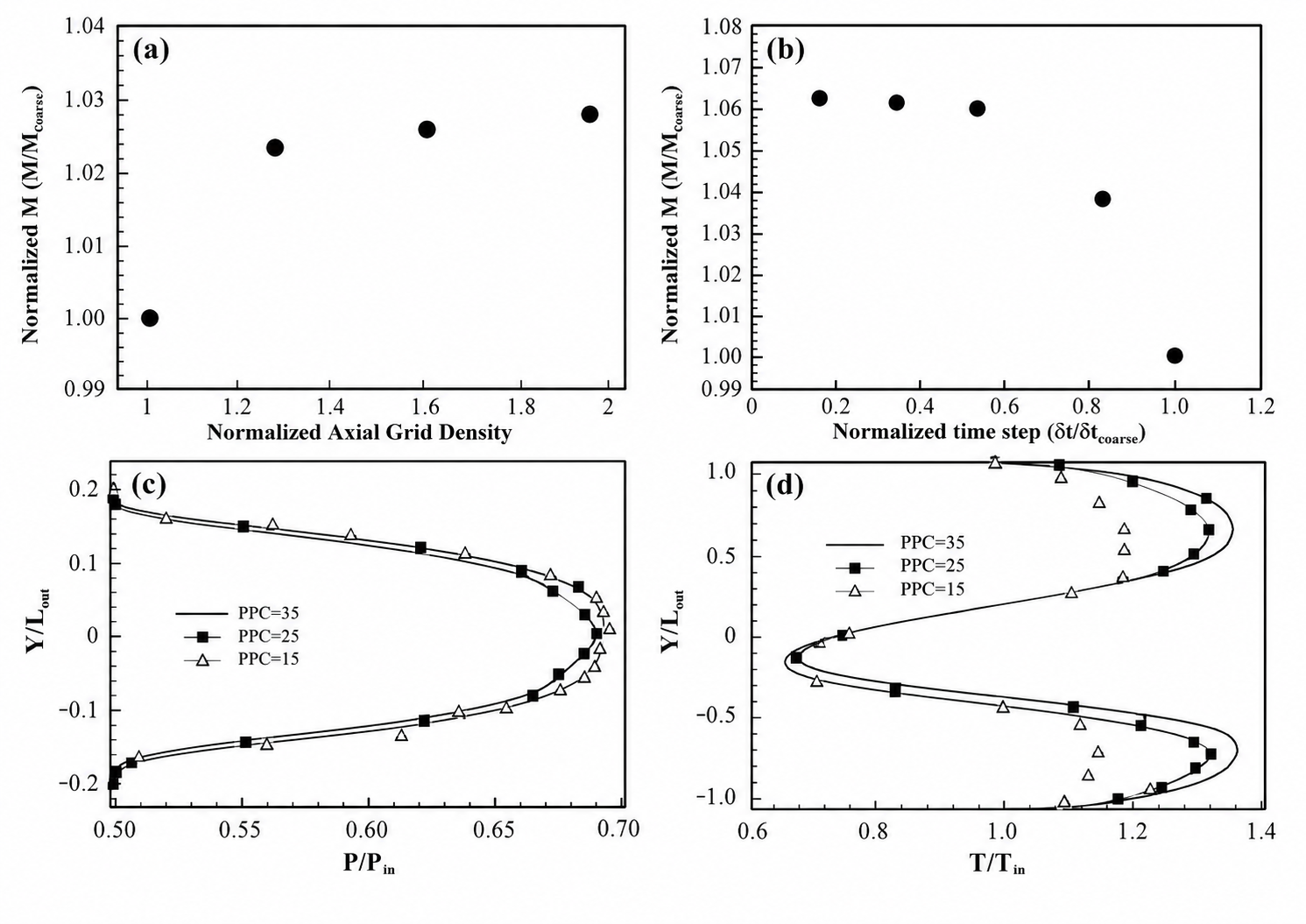}
\caption{Combined numerical verification of grid density, time step, and particle-per-cell sensitivity. (a) Grid-independence study for the Heating~2 case, $Q_w=2.5\times10^4~{\rm W\,m^{-2}}$, based on the area-weighted average throat Mach number normalized by the coarsest-grid value, $M/M_{\rm coarse}$. (b) Time-step sensitivity study for the strongest heating case, $Q_w=7.5\times10^4~{\rm W\,m^{-2}}$, based on the average exit Mach number normalized by the coarsest-time-step value; the result supports the production choice $\Delta t=4\times10^{-10}~{\rm s}$. (c) Particle-number sensitivity for the Heating~1 case, $Q_w=1.0\times10^4~{\rm W\,m^{-2}}$, based on the normalized throat pressure profile, $P/P_{\rm in}$, for 15, 25, and 35 particles per cell. (d) Particle-number sensitivity for the same case based on the normalized outlet temperature profile, $T/T_{\rm in}$. The agreement between the 25- and 35-particle cases supports the use of 25 particles per cell in the final simulations.}
\label{fig:verification}
\end{figure}

Numerical verification is summarized in Fig.~\ref{fig:verification}. The purpose of this test is to ensure that the numerical settings used in the production simulations are not responsible for the observed heat-flux trends. Three independent checks are included: grid refinement, time-step sensitivity, and particle-number sensitivity. The final numerical parameters reported in Table~\ref{tab:dsmc_parameters} were selected based on these tests.

Figure~\ref{fig:verification}(a) shows the grid-independence study. This test was performed for the Heating~2 case, $Q_w=2.5\times10^4~{\rm W\,m^{-2}}$, because this case already contains strong wall thermal forcing while remaining below the extreme heating limit. The monitored quantity is the area-weighted average Mach number at the nozzle throat, normalized by the value obtained on the coarsest grid, $M/M_{\rm coarse}$. The throat region was selected because it is highly sensitive to acceleration, viscous blockage, and local grid resolution. The normalized Mach number changes noticeably from the coarsest grid to the intermediate grids, but the variation becomes very small between the two finest grids. Therefore, the production grid listed in Table~\ref{tab:dsmc_parameters}, with $25\times30$ cells in the converging block, $75\times30$ cells in the diverging block, and $30\times40$ cells in the downstream buffer block, was adopted for the heat-flux simulations.

Figure~\ref{fig:verification}(b) shows the time-step sensitivity study. This test was performed for the strongest heating case, $Q_w=7.5\times10^4~{\rm W\,m^{-2}}$, because it produces the largest wall-temperature rise, strongest molecular-speed variation, and most restrictive temporal resolution requirement among the simulated cases. The monitored quantity is the average Mach number at the nozzle exit plane, normalized by the value obtained using the coarsest time step, $M/M_{\rm coarse}$. The response changes significantly for coarse time steps but approaches an asymptotic value as the time step is reduced. The time step $\Delta t=4\times10^{-10}~{\rm s}$ gives a response close to the finer time-step results while maintaining reasonable computational cost; it was therefore used in the production simulations.

Figures~\ref{fig:verification}(c) and \ref{fig:verification}(d) show the particle-number sensitivity study. This test was performed for the Heating~1 case, $Q_w=1.0\times10^4~{\rm W\,m^{-2}}$. Panel~(c) compares the normalized static-pressure profile, $P/P_{\rm in}$, at the nozzle throat for 15, 25, and 35 particles per cell. Panel~(d) compares the normalized temperature profile, $T/T_{\rm in}$, at the nozzle outlet for the same particle populations. The 15-particle case exhibits visible statistical scatter and profile deviation, whereas the 25- and 35-particle cases nearly collapse for both pressure and temperature. Consequently, 25 particles per cell were selected for the production runs as a balance between statistical convergence and computational cost.

\section{Field-derived diagnostics}
\label{sec:configuration}

\subsection{Aerodynamic and rarefaction diagnostics}

The central diagnostic introduced in this paper is the effective mass-flux thickness. On the simulated half-domain, the axial mass-flux density is
\begin{equation}
j_m(x,y)=\rho(x,y)u(x,y).
\end{equation}
The mass-flux thickness is defined as
\begin{equation}
h_m(x)=
\frac{\displaystyle\int_{y_w}^{y_s} j_m(x,y)\,dy}
{\displaystyle \max_{y\in[y_w,y_s]}j_m(x,y)},
\label{eq:hm}
\end{equation}
where $y_w$ is the wall and $y_s$ is the symmetry plane. The local geometric half-height is
\begin{equation}
h_{\rm geo}(x)=|y_s-y_w|.
\end{equation}
The normalized mass-flux thickness and the effective blockage are
\begin{equation}
\beta_m(x)=\frac{h_m(x)}{h_{\rm geo}(x)},\qquad
B_m(x)=1-\beta_m(x).
\label{eq:blockage}
\end{equation}
When $\beta_m$ is close to unity, most of the geometric passage carries mass flux effectively. When $\beta_m$ decreases, a larger portion of the passage contains low-momentum gas and the effective mass-carrying core contracts. This definition is more appropriate than a conventional continuum displacement thickness for the present DSMC data because it is tied directly to the sampled mass-flux distribution.

To visualize compression and expansion, we use the signed numerical-schlieren field
\begin{equation}
\chi(x,y)=\frac{\partial(\rho/\rho_0)}{\partial(x/L)}.
\label{eq:schlieren}
\end{equation}
This quantity retains the sign of streamwise compression and expansion. The unsigned magnitude $|\nabla\rho|$ is useful for detecting high-gradient regions, but it merges compression, expansion, wall gradients and corner features into a single positive field. The signed field is therefore more useful for interpreting the shock-cell-like compression signature.

A gradient-length Knudsen-number diagnostic is also used. For a variable $\phi$, the gradient-length local Knudsen number is
\begin{equation}
Kn_{\rm GLL,\phi}=\lambda\frac{|\nabla\phi|}{|\phi|+\epsilon_\phi},
\end{equation}
Here $\epsilon_\phi$ is a small field-dependent numerical floor introduced only to avoid division by zero when the normalized field magnitude becomes locally very small. It does not affect the high-gradient regions discussed below. The composite indicator is
\begin{equation}
Kn_{\rm GLL}=\max_{\phi\in\{\rho,T,p,Ma\}}Kn_{\rm GLL,\phi}.
\end{equation}
This diagnostic highlights local continuum breakdown driven by short gradient lengths, not only by the global molecular mean free path.

Finally, the total thrust and specific impulse are computed from exit-plane integrals,
\begin{equation}
\dot{m}=\int_A \rho u_x\,dA,
\end{equation}
\begin{equation}
F_t = \int_A \rho u_x^2\,dA+\int_A (p-p_{\rm amb})\,dA,
\end{equation}
\begin{equation}
I_{sp}=\frac{F_t}{\dot{m}g_0}.
\label{eq:isp}
\end{equation}
The same exit plane is used for $\dot{m}$, $F_t$ and $I_{sp}$ so that the integrated performance metrics remain internally consistent.

\subsection{Heat-transfer and viscous--thermal scaling diagnostics}

Because the simulations are rarefied and locally non-equilibrium, we do not compute a field heat-flux vector from Fourier's law and we do not interpret a continuum wall-normal temperature gradient as the primary heat-transfer result. The prescribed and sampled gas--surface heat flux is the reliable heat-transfer quantity. To compare the wall-to-bulk thermal response among cases, we define a local apparent heat-flux-based Nusselt-type response using a temperature-dependent conductivity,
\begin{equation}
Nu_q^{\rm loc}(x)=
\frac{Q_wD_h(x)}{k_f(x)\left[T_w(x)-T_b(x)\right]}.
\label{eq:nuq}
\end{equation}
Here $D_h(x)=2h_{\rm geo}(x)$ is the local planar hydraulic height and $T_b(x)$ is the mass-flux-weighted bulk gas temperature,
\begin{equation}
T_b(x)=\frac{\displaystyle\int_{y_w}^{y_s}\rho u T\,dy}{\displaystyle\int_{y_w}^{y_s}\rho u\,dy}.
\label{eq:tb}
\end{equation}
The conductivity is evaluated at a wall--bulk film temperature,
\begin{equation}
T_f(x)=\frac{T_w(x)+T_b(x)}{2},
\qquad
\mu_f(x)=\mu_{\rm ref}\left[\frac{T_f(x)}{T_{\rm ref}}\right]^\omega,
\label{eq:mufilm}
\end{equation}
and is converted from the VHS viscosity using a nitrogen Eucken factor consistent with the kinetic treatment of non-vibrating polyatomic gases,
\begin{equation}
k_f(x)=f_{{\rm eu},N_2}\frac{f_{\rm dof}}{2}\frac{k_B}{m}\mu_f(x).
\label{eq:kvhs}
\end{equation}
For the present nitrogen simulations, translational and rotational modes are active while vibrational excitation is neglected; hence $f_{\rm dof}=5$. Following the polyatomic-gas formulation and the nitrogen Eucken-factor values reported by Wu and co-workers \citep{wu2015polyatomic,wu2020eucken}, we use $f_{{\rm eu},N_2}=1.96$, so that $k_f=4.90(k_B/m)\mu_f$. The use of a film temperature is consistent with the fact that $Nu_q^{\rm loc}$ compares a wall heat flux with a wall--bulk temperature difference. 

The sign of $Nu_q^{\rm loc}$ is retained. When $T_w-T_b$ approaches zero, the imposed-flux-to-temperature-difference ratio becomes mathematically singular. These locations are not numerical errors and are not removed from the physical interpretation; they mark the wall--bulk temperature crossings discussed later. For comparative plots of finite Nusselt levels, we use a validity mask $|T_w-T_b|>0.05T_0$, while the raw signed response is reported in ~\ref{app:raw_nusselt}. The mask is therefore a conditioning mask for the ratio definition, not a smoothing or denoising operation.

A complementary Brinkman-type measure is introduced to compare the local viscous momentum-transport scale with the imposed thermal forcing. Because the VHS molecular model gives a temperature-dependent viscosity, the viscosity is evaluated from the mass-flux-weighted bulk temperature rather than held fixed:
\begin{equation}
\mu_b(x)=\mu_{\rm ref}\left[\frac{T_b(x)}{T_{\rm ref}}\right]^\omega,
\label{eq:muvhs}
\end{equation}
where $T_{\rm ref}=273~{\rm K}$ and $\omega=0.74$ for nitrogen in the present VHS model. The local Brinkman-type imposed-flux ratio is then defined as
\begin{equation}
Br_q^{\rm loc}(x)=\frac{\mu_b(x)U_b^2(x)}{|Q_w|D_h(x)+\epsilon_Q},
\label{eq:brq}
\end{equation}
where $U_b(x)$ is the mass-flux-weighted axial velocity. Small $Br_q^{\rm loc}$ indicates that the imposed wall heat flux is large relative to the local viscous momentum-transport scale, while large $Br_q^{\rm loc}$ indicates a more momentum-dominated thermal response. This definition is consistent with the temperature-dependent transport scaling used by the collision model and avoids comparing strongly heated and cooled cases with a fixed viscosity.

\section{Results and discussion}
\label{sec:results}

\subsection{Coupled wall and bulk temperature response}

\begin{figure}[!t]
\centering
\includegraphics[
  width=0.80\textwidth,
  height=0.34\textheight,
  keepaspectratio
]{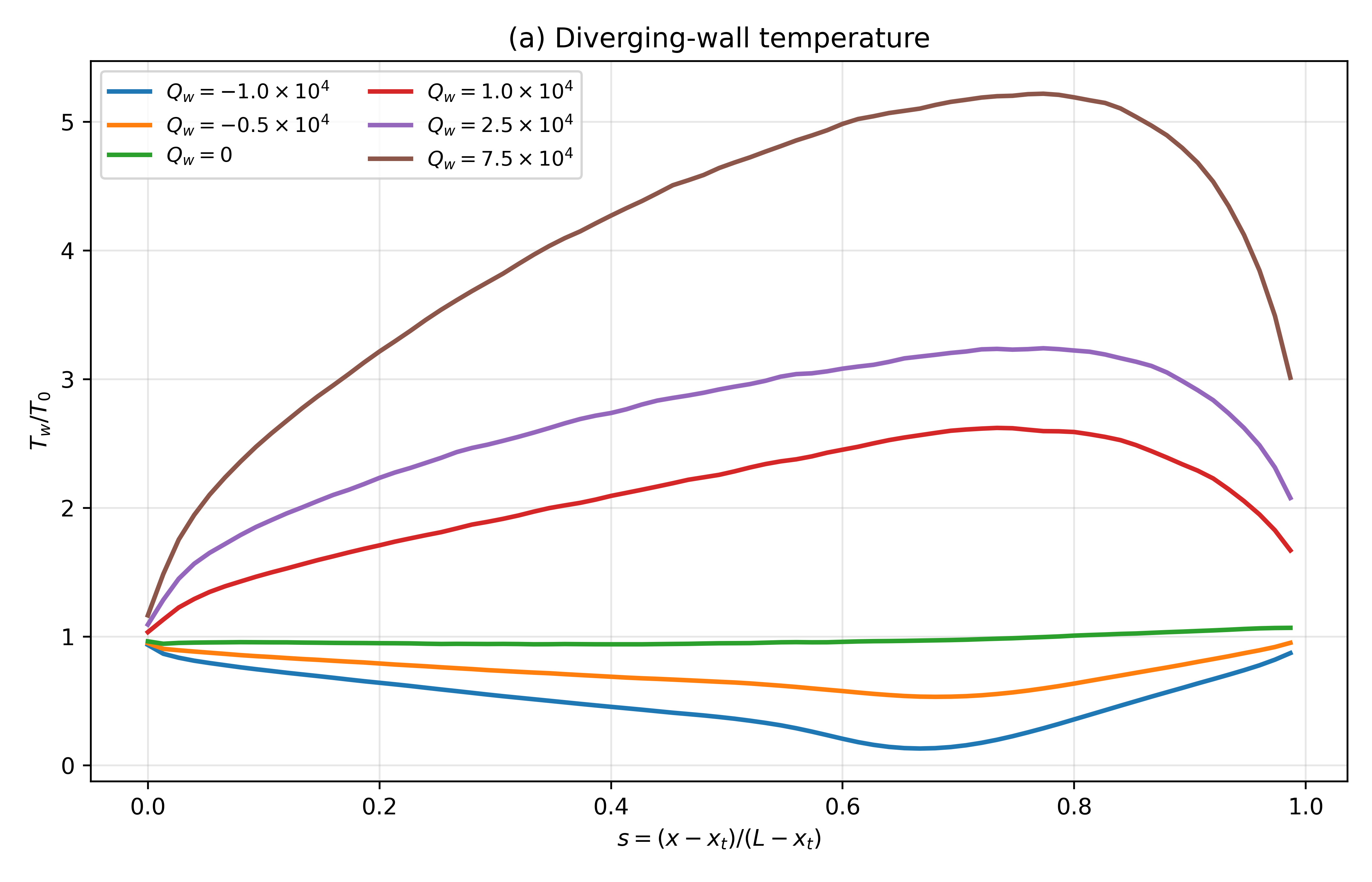}

\vspace{0.4em}

\includegraphics[
  width=0.80\textwidth,
  height=0.34\textheight,
  keepaspectratio
]{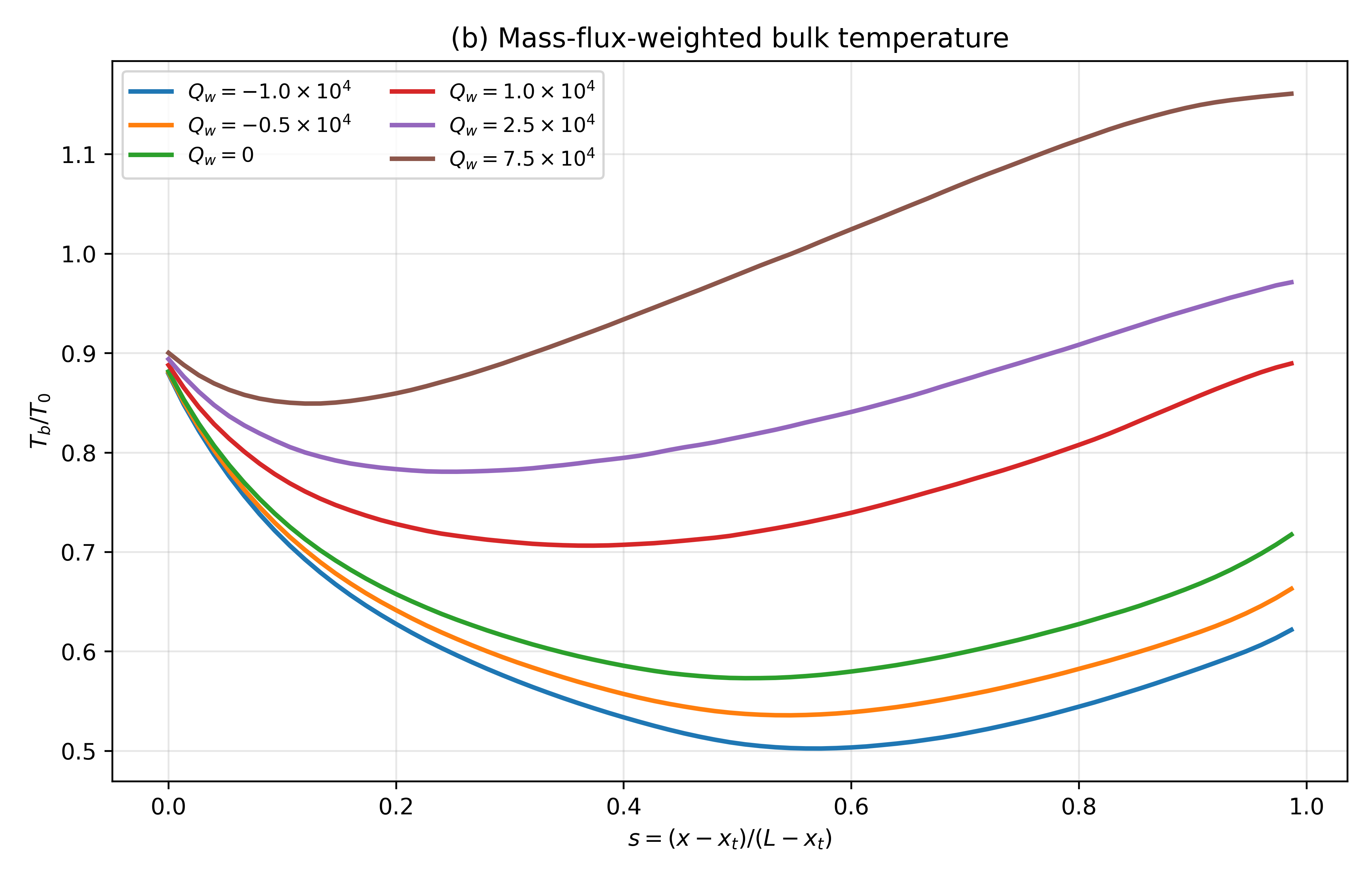}

\caption{Coupled wall--bulk temperature response on the heated/cooled diverging wall. 
(a) Normalized diverging-wall temperature $T_w/T_0$. Cooling drives the wall below the inlet temperature, the adiabatic wall remains close to unity, and strong heating increases $T_w/T_0$ to values above five near the exit. 
(b) Mass-flux-weighted bulk temperature $T_b/T_0$. The bulk gas initially cools by expansion for all cases, but sufficiently strong heating reverses the downstream trend and raises the exit bulk temperature. The contrast between panels (a) and (b) shows that imposed wall heat flux produces wall--bulk stratification rather than a spatially uniform gas-temperature shift.}
\label{fig:walltemp}
\end{figure}

Figure~\ref{fig:walltemp} presents the coupled wall--bulk thermal diagnostic. The wall-temperature response in Fig.~\ref{fig:walltemp}(a) confirms that the feedback heat-flux boundary condition converges to physically distinct wall states. Upstream of the throat, the wall temperature remains close to the reference value for all cases because no thermal forcing is imposed there. Downstream of the throat, the curves separate according to $Q_w$: cooling lowers the wall temperature, the adiabatic case remains close to unity, and heating produces a strong monotonic increase along the diverging wall.

The bulk gas-temperature field in Fig.~\ref{fig:walltemp}(b) is the important new information. Heating does not remain a wall-only boundary effect. It creates a hot near-wall layer that grows downstream and progressively occupies a larger fraction of the diverging passage. The central core remains much cooler than the wall in the strongest heating case, which means that the imposed heat flux produces strong wall-to-core thermal stratification. Cooling produces the opposite behavior: it maintains a colder diverging section and suppresses the near-wall thermal layer. This wall--bulk separation is the physical origin of the blockage and propulsion trends discussed below.

\subsection{Near-wall tangential slip response}

\begin{figure}[H]
\centering
\includegraphics[width=0.94\textwidth]{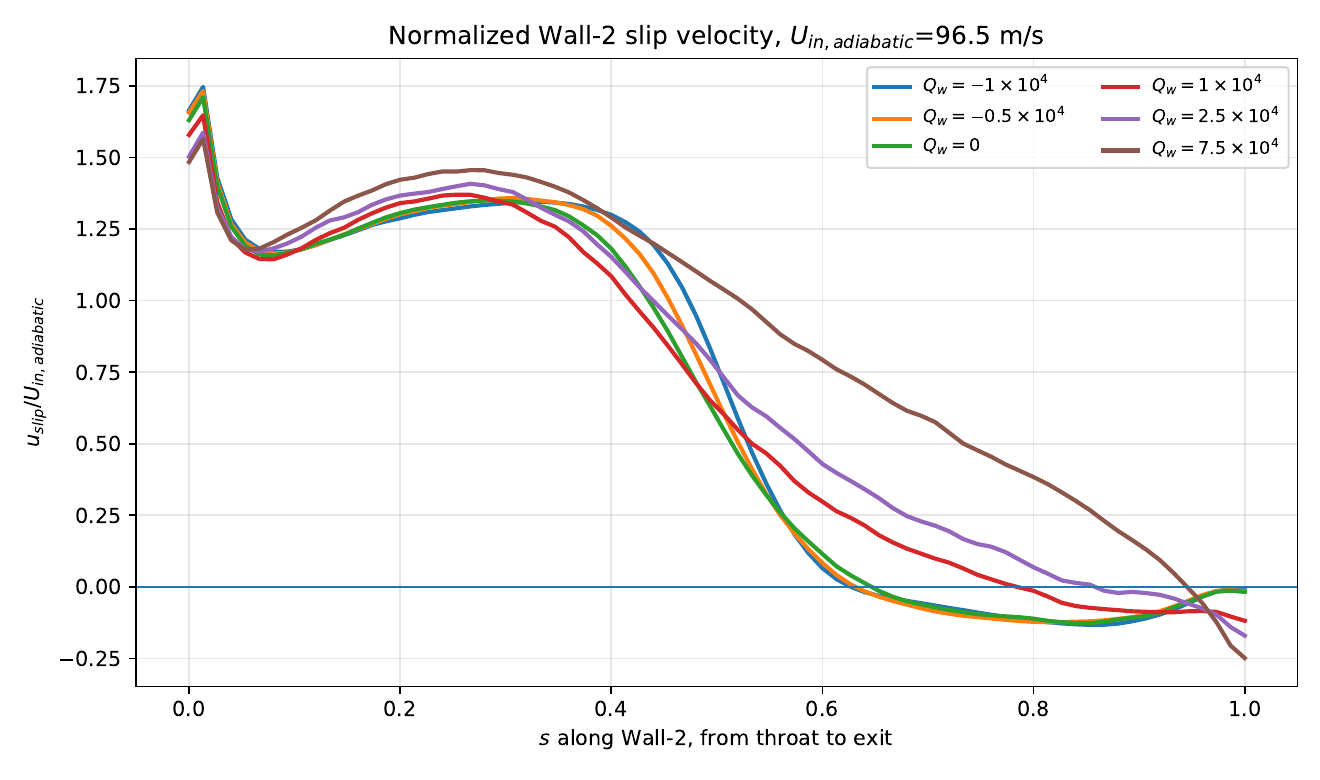}
\caption{Normalized near-wall tangential slip response on Wall-2. The plotted quantity is computed from the first gas layer adjacent to the thermally forced wall and projected onto the local wall-tangent direction, $u_{\rm slip}=\bm{u}_{j=1}\cdot\bm{t}_w$, then normalized by the inlet velocity of the adiabatic case. Strong heating maintains a larger positive near-wall tangential response over a longer portion of the diverging wall, consistent with the growth of a wall-driven viscous--thermal layer and the contraction of the effective mass-carrying core.}
\label{fig:slip}
\end{figure}

To connect the wall thermal forcing to the momentum response at the gas--surface interface, Fig.~\ref{fig:slip} reports the near-wall tangential velocity on Wall-2. The quantity is evaluated from the first gas layer adjacent to the wall and projected onto the local wall-tangent direction,
\begin{equation}
u_{\rm slip}(s)=\bm{u}_{j=1}(s)\cdot\bm{t}_w(s),
\label{eq:slip}
\end{equation}
where $s=(x-x_t)/(L-x_t)$ is the normalized coordinate along the diverging wall. Because the wall is stationary, this projection is the tangential velocity mismatch between the gas sampled in the first near-wall layer and the wall. It should therefore be interpreted as a DSMC near-wall slip-response diagnostic, not as a continuum Maxwell-slip coefficient obtained from a fitted velocity gradient.

The near-wall tangential response in Fig.~\ref{fig:slip} is interpreted using Maxwell's general slip framework. As emphasized by Maxwell and later by Lockerby et al.~\citep{maxwell1879,lockerby2004velocity}, the fundamental boundary condition relates tangential slip to gas--surface tangential momentum and energy exchange, rather than to a velocity gradient alone. In symbolic stress/heat-flux form,
\begin{equation}
u_{\rm slip}
\sim
A_\sigma \frac{\lambda}{\mu}\tau_{nt}
+
A_q\frac{q_t}{p},
\label{eq:maxwell_general_slip}
\end{equation}
where $\tau_{nt}$ is the tangential shear stress at the wall, $q_t=\mathbf{q}\cdot\mathbf{t}_w$ is the tangential component of the heat-flux vector along the wall, $\lambda$ is the local mean free path, $\mu$ is the local viscosity, $p$ is the local pressure, and $\mathbf{t}_w$ is the local wall-tangent direction. The constants $A_\sigma$ and $A_q$ depend on the gas--surface accommodation model and on the sign convention used for $\tau_{nt}$ and $q_t$. This form is used only for interpretation; it emphasizes the correct physical ordering that the slip response is controlled by tangential momentum transfer and tangential heat transport at the gas--surface interface.

To connect Eq.~\eqref{eq:maxwell_general_slip} to the present DSMC field data, we introduce continuum-style diagnostic approximations for the two contributions. The shear-stress term is approximated locally as
\begin{equation}
\tau_{nt}
\simeq
\mu
\left.\frac{\partial u_t}{\partial n}\right|_w ,
\label{eq:shear_stress_proxy}
\end{equation}
where $u_t=\mathbf{u}\cdot\mathbf{t}_w$ is the gas velocity tangent to Wall-2 and $n$ is the wall-normal coordinate. Substitution into the first term of Eq.~\eqref{eq:maxwell_general_slip} gives the usual first-order shear-slip scaling,
\begin{equation}
u_{\rm slip}^{(\tau)}
\sim
A_\sigma \lambda
\left.\frac{\partial u_t}{\partial n}\right|_w .
\label{eq:shear_slip_scaling}
\end{equation}
Thus, the shear-driven part depends not only on the near-wall tangential velocity gradient, but also on the local mean free path.

The thermal-creep term is connected to the tangential heat flux through Fourier's approximation,
\begin{equation}
q_t
\simeq
-k_f
\left.\frac{\partial T}{\partial s}\right|_w ,
\label{eq:fourier_tangential_heatflux}
\end{equation}
where $s$ is the wall-tangent coordinate and $k_f$ is the local film-temperature thermal conductivity. For an ideal gas with temperature-dependent viscosity and Eucken-type conductivity, this term is proportional to
\begin{equation}
u_{\rm slip}^{(T)}
\sim
C_T
\frac{\mu}{\rho T}
\left.\frac{\partial T_w}{\partial s}\right|_w .
\label{eq:thermal_creep_scaling}
\end{equation}
The local mean free path is also embedded in this thermal-creep scaling because
\begin{equation}
\lambda =
\frac{\mu}{p}\sqrt{\frac{\pi R T}{2}},
\qquad
p=\rho R T,
\label{eq:vhs_mfp}
\end{equation}
with $\mu=\mu_{\rm ref}(T/T_{\rm ref})^\omega$ for the VHS model. Hence, $\mu/(\rho T)=R\mu/p$ is proportional to the local mean free path. The thermal-creep contribution therefore depends on both the wall-tangential temperature gradient and the local rarefaction/transport scale.

The DSMC data fields are used to construct two diagnostic proxies. The shear-driven proxy is
\begin{equation}
\mathcal{S}(s)
=
\lambda
\frac{\Delta u_t}{\Delta n},
\label{eq:shear_proxy_slip}
\end{equation}
where the velocities in the first two gas layers adjacent to Wall-2 are projected onto the local wall tangent before differencing in the wall-normal direction. The thermal-creep proxy is
\begin{equation}
\mathcal{T}(s)
=
\frac{\mu}{\rho T}
\frac{dT_w}{ds},
\label{eq:thermal_creep_proxy}
\end{equation}
with the sign convention chosen so that a positive wall-temperature gradient along the positive wall-tangent direction corresponds to a positive thermal-creep contribution. These quantities are not exact slip coefficients and are not substitutes for direct wall-stress or wall-heat-flux sampling. They are used only to check whether the sign and streamwise variation of the measured near-wall tangential response are consistent with Maxwell's shear-slip and thermal-creep mechanisms.

The mean-free-path variation is important for this interpretation. Using the VHS viscosity law and the local near-wall pressure and temperature, the computed $\lambda$ remains relatively small in the cooling cases and increases strongly with heating. In the middle portion of Wall-2, the median near-wall mean free path is approximately $0.29$--$0.41~\mu{\rm m}$ in the two cooling cases, about $0.55~\mu{\rm m}$ in the adiabatic case, and increases to approximately $1.2$, $1.5$, and $2.3~\mu{\rm m}$ for $Q_w=1.0\times10^4$, $2.5\times10^4$, and $7.5\times10^4~{\rm W\,m^{-2}}$, respectively. Heating therefore amplifies the effective slip length as well as modifying the velocity and temperature gradients.

The shear proxy explains the large positive response immediately downstream of the throat. In this region, the gas accelerates rapidly along the diverging wall, and $\mathcal{S}(s)$ is positive for all thermal cases. This produces the strong positive $u_{\rm slip}/U_{in,ad}$ observed at small $s$. Farther downstream, the thermal cases separate because the imposed heat flux changes three coupled quantities: the near-wall tangential velocity gradient, the wall-tangential temperature gradient, and the local mean free path. In the cooling cases, the wall temperature decreases along the upstream and middle portions of Wall-2, so $dT_w/ds<0$ over much of this region. The thermal-creep proxy $\mathcal{T}(s)$ is therefore negative and opposes the positive shear-driven slip. Cooling also lowers the local temperature and viscosity and keeps $\lambda$ smaller, reducing the effective slip length. Consequently, the near-wall tangential response decays rapidly and crosses zero around the middle-to-late part of Wall-2.

Heating produces the opposite upstream and mid-wall behavior. The heated wall raises the near-wall molecular energy, increases the local viscosity and mean free path, and broadens the wall-driven viscous--thermal layer. The shear proxy remains positive over a longer streamwise distance, and the larger $\lambda$ increases the effective shear-slip scaling. In addition, for much of the upstream and middle diverging wall, $dT_w/ds>0$, so the thermal-creep proxy has the same sign as the shear-driven response. The two mechanisms therefore reinforce each other: the shear term reflects the sustained tangential momentum imbalance between the gas and the stationary wall, while the thermal-creep term reflects tangential migration associated with the wall-tangential heat flux. This is why the heating cases, especially $Q_w=7.5\times10^4~{\rm W\,m^{-2}}$, maintain a positive $u_{\rm slip}/U_{in,ad}$ over a much longer portion of Wall-2 than the cooling and adiabatic cases.

The downstream decrease and weak sign reversal near the outlet are also consistent with the same decomposition. Near the outlet, the wall-temperature profiles bend downward in the heated cases, so $dT_w/ds$ becomes negative and the thermal-creep proxy reverses sign. For the strongest heating case, the shear proxy remains positive over most of Wall-2 and $\lambda$ remains large, but the downstream thermal-creep proxy becomes negative and reduces the net tangential response near the exit. The sign reversal should therefore not be interpreted as a classical separated recirculation zone. It is better understood as the combined effect of outlet/compression-zone reorientation, weakening of the local tangential momentum gradient, and reversal of the wall-tangential heat-flux contribution.

The analysis based on Eqs.~\eqref{eq:shear_proxy_slip} and \eqref{eq:thermal_creep_proxy} should be viewed as a first-order interpretive estimate. The local Knudsen-number variations and the finite Knudsen layer mean that the continuum substitutions in Eqs.~\eqref{eq:shear_stress_proxy} and \eqref{eq:fourier_tangential_heatflux} are not exact wall boundary relations for the DSMC solution. Nevertheless, they provide the correct physical decomposition needed to interpret Fig.~\ref{fig:slip}: wall heating increases the effective slip length, sustains a positive shear-driven contribution, and adds a reinforcing thermal-creep contribution over much of the diverging wall. This combined response expands the wall-affected viscous--thermal layer and is consistent with the contraction of the effective mass-carrying core observed in the blockage metric.

\subsection{Signed numerical schlieren and finite compression structures}

\begin{figure}[H]
\centering
\includegraphics[width=1.1\textwidth]{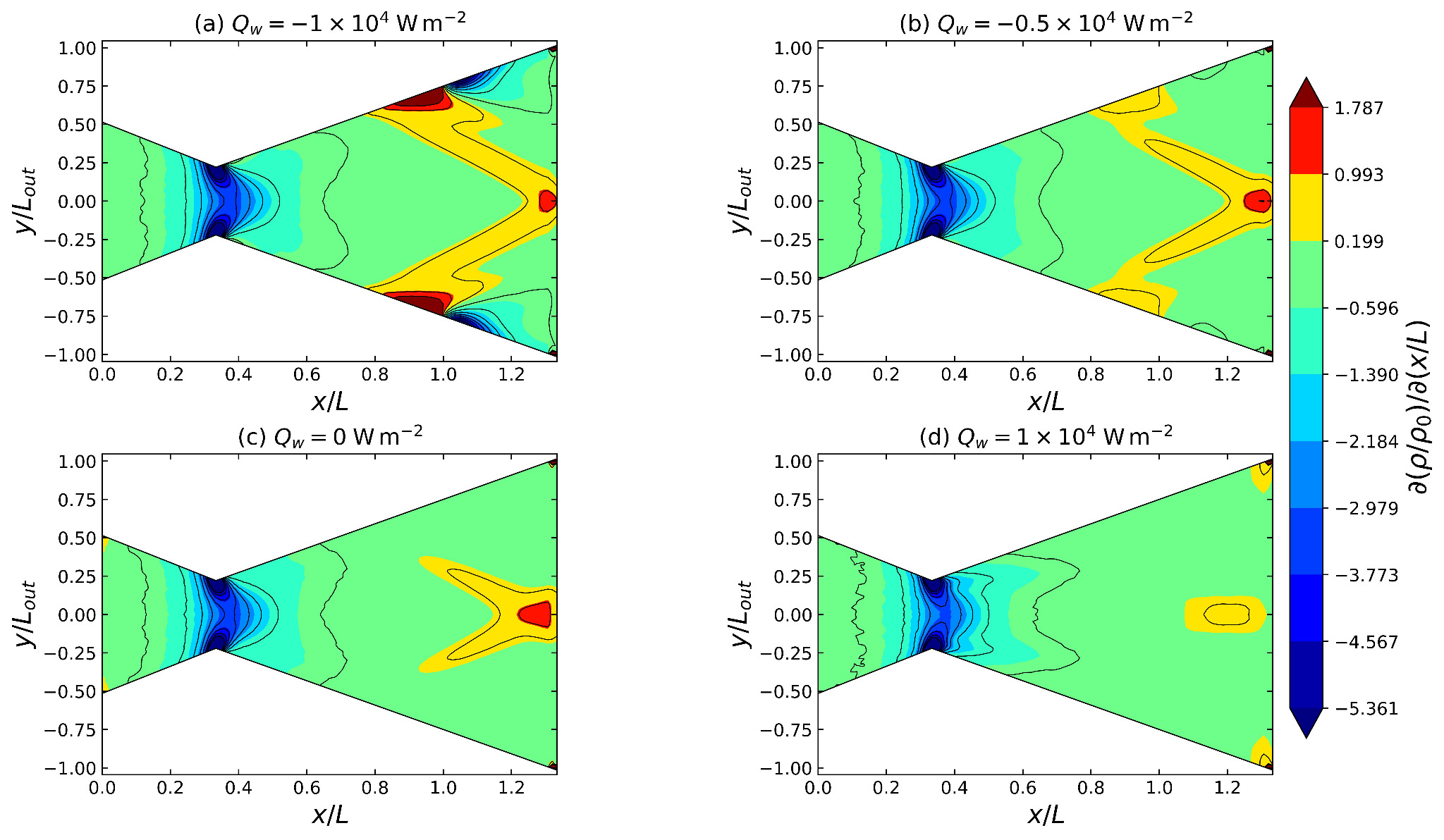}
\caption{Signed numerical-schlieren field $\partial(\rho/\rho_0)/\partial(x/L)$ for representative heat-flux cases. Cooling produces a sharper shock-cell-like compression signature, whereas heating weakens and spreads the compression into a broader viscous--thermal zone.}
\label{fig:schlieren}
\end{figure}

The signed numerical-schlieren field in Fig.~\ref{fig:schlieren} shows the streamwise density-gradient structure,
\[
\chi=\frac{\partial(\rho/\rho_0)}{\partial(x/L)} .
\]
The sign is important. Blue regions correspond to a negative streamwise density gradient, i.e. streamwise expansion or density depletion, whereas yellow--red regions correspond to a positive streamwise density gradient, i.e. compression. The strong blue region around the throat is therefore not a shock signature. It is the local expansion/acceleration region generated as the flow passes through the minimum-area section and enters the diverging passage. In this region the gas accelerates rapidly, the static pressure and density drop over a short axial distance, and the finite-thickness rarefied expansion layer appears as a negative signed schlieren band. The blue throat feature is strongest where the area change and streamwise acceleration are largest, and it persists in all thermal cases because it is primarily controlled by the nozzle geometry and pressure ratio.

The downstream positive-gradient structure has a different origin. The yellow--red region near the outlet and downstream buffer corresponds to the internal compression or shock-cell-like response produced by the finite back pressure. In the cooling cases, the wall remains cold, the near-wall viscosity and thermal layer are comparatively weaker, and a larger fraction of the geometric passage remains available to the high-speed core. The density adjustment therefore occurs over a shorter axial distance, producing a sharper positive signed-schlieren ridge. This is why the cooling cases show a more distinct compression pattern near the outlet and near the wall-interaction regions.

Heating changes this picture in several coupled ways. First, wall heating raises the near-wall molecular energy and increases the local viscosity and mean free path. This thickens the wall-driven viscous--thermal layer and reduces the effective mass-carrying core, as quantified later by the decrease of $\beta_m$ and the increase of effective blockage. Second, because part of the imposed wall energy is converted into internal energy of the near-wall gas, the axial acceleration of the core is weakened and the peak Mach number is reduced. Third, the compression is no longer concentrated into a narrow shock-cell-like ridge. Instead, the density rise is distributed over a broader region because molecular transport and wall-layer growth smear the streamwise density adjustment. Thus, heating does not make compression disappear; it changes the compression from a sharp, localized signed-density-gradient feature into a broader viscous--thermal compression zone.

This interpretation is consistent with the other diagnostics in the paper. The wall--bulk temperature profiles show that heating creates strong thermal stratification rather than a uniform gas-temperature shift. The slip and viscosity diagnostics show that the heated wall sustains a thicker wall-affected layer. The blockage metric shows that this layer contracts the effective mass-carrying passage. The mass-flow and discharge-coefficient trends show the corresponding throughput penalty, while the thrust and specific impulse show that thermal augmentation can still increase impulse despite the increased blockage. Therefore, the smoother field in the strongest heating case should not be described as a complete disappearance of the internal compression. A more precise statement is that the distinct shock-cell-like compression ridge is weakened, shifted, and merged into a broader viscous--thermal compression region.

\subsection{Gradient-length rarefaction and redistribution of non-equilibrium}

\begin{figure}[H]
\centering
\includegraphics[width=1.1\textwidth]{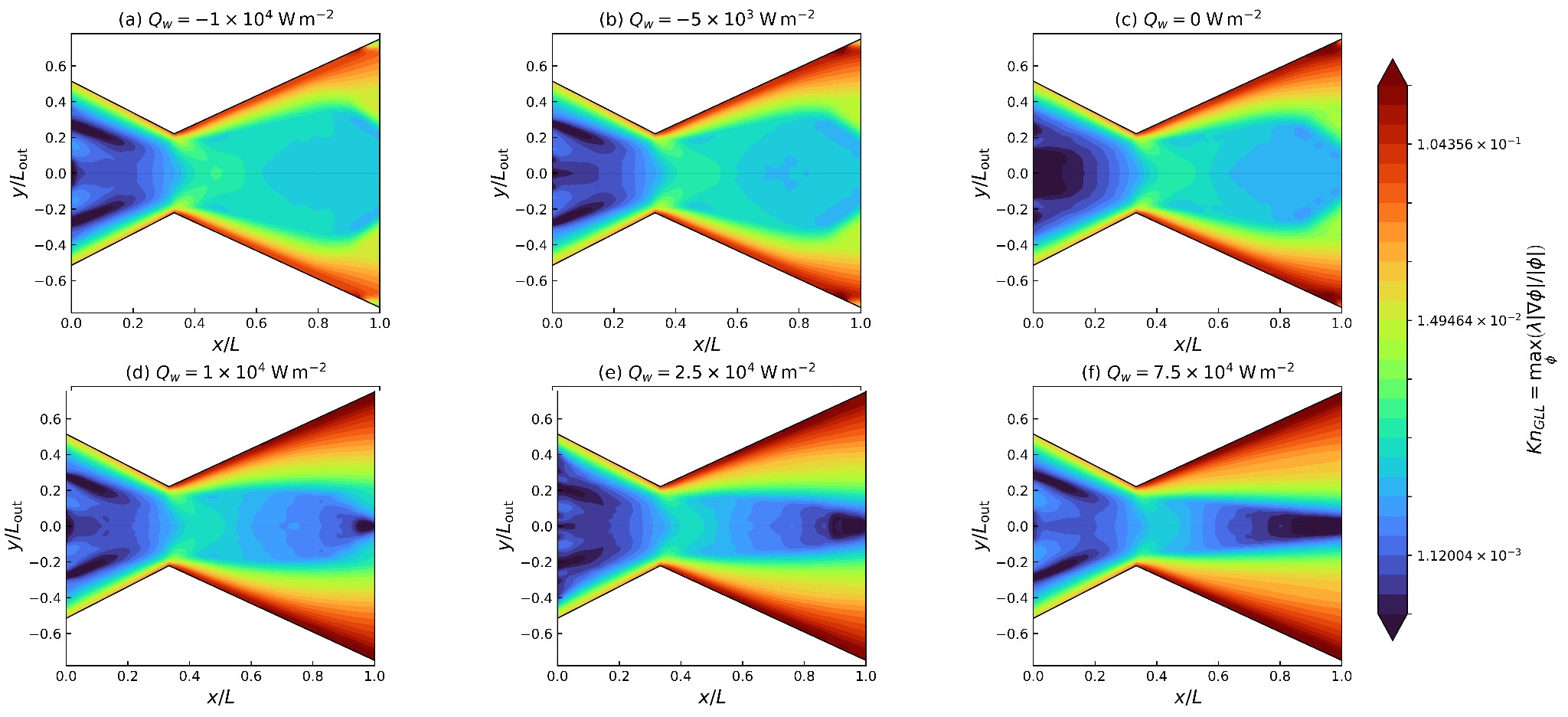}
\caption{Composite gradient-length Knudsen-number field for the heat-flux sweep. The plotted quantity is \(Kn_{\rm GLL}=\max_{\phi\in\{\rho,T,p,Ma\}}\lambda|\nabla\phi|/(|\phi|+\epsilon_\phi)\). Heating does not simply move a compact compression feature; it expands a wall-driven viscous--thermal region of short-gradient-length transport and confines the lower-gradient mass-carrying core toward the nozzle centerline.}
\label{fig:kngll}
\end{figure}

Figure~\ref{fig:kngll} shows the composite gradient-length Knudsen-number field. As defined in Section~\ref{sec:configuration}, this diagnostic is not the conventional global Knudsen number. For each field variable, 
\[
Kn_{{\rm GLL},\phi}=\lambda\frac{|\nabla\phi|}{|\phi|+\epsilon_\phi},
\qquad 
\phi\in\{\rho,T,p,Ma\},
\]
and the plotted field is the composite maximum,
\[
Kn_{\rm GLL}=\max_{\phi\in\{\rho,T,p,Ma\}} Kn_{{\rm GLL},\phi}.
\]
Thus, the figure identifies locations where the molecular transport length is no longer small compared with the local gradient length of density, temperature, pressure, or Mach number. It should therefore be interpreted as a field-level indicator of local non-equilibrium and short-gradient-length transport, not as a map of the global rarefaction level alone.

The distribution in Fig.~\ref{fig:kngll} is controlled by two factors: the local mean free path, $\lambda$, and the inverse gradient length, $|\nabla\phi|/(|\phi|+\epsilon_\phi)$. Regions of large $Kn_{\rm GLL}$ can therefore arise either because the gas is locally more rarefied or because one of the macroscopic fields varies over a short distance. The max operation also means that the dominant variable need not be the same everywhere. Near the throat and in the internal compression region, the pressure, density, and Mach-number gradients contribute strongly. Along the heated diverging wall, however, the temperature and pressure/density gradients associated with the wall-driven viscous--thermal layer become the dominant source of the large composite response. This is why the strongest $Kn_{\rm GLL}$ levels appear close to the solid boundaries and not only in the nominal shock-cell region.

The high values near the diverging wall should not be interpreted as a numerical artifact. They are a direct consequence of the imposed heat flux. Heating raises the wall temperature, increases the local viscosity and mean free path, and creates a strong wall-normal thermal and density stratification. At the same time, the thermally modified near-wall layer carries less effective axial mass flux than the core, as later quantified by the reduction of $\beta_m$ and the increase of $1-\beta_m$. In this sense, Fig.~\ref{fig:kngll} provides the rarefaction-side counterpart of the blockage analysis: the same wall-driven layer that reduces the effective mass-carrying thickness also occupies a larger fraction of the diverging passage with short-gradient-length, non-equilibrium transport.

The blue/green central region in the heated cases should therefore be read carefully. It does not mean that the near-wall flow has disappeared, nor does it imply classical separation. Rather, it shows that the lowest-gradient part of the flow is increasingly confined to a narrower central core, while the wall-adjacent region is dominated by viscous--thermal gradients. This interpretation is consistent with the mass-flux-thickness diagnostic: heating does not remove gas from the near-wall region, but it shifts the effective mass-carrying core away from the heated wall and makes the nozzle behave aerodynamically as if the available passage were smaller than the geometric passage.

This also explains why the compression signature is less visually dominant in Fig.~\ref{fig:kngll} than in the signed numerical-schlieren field. The signed schlieren diagnostic isolates the streamwise density gradient,
\[
\chi=\frac{\partial(\rho/\rho_0)}{\partial(x/L)},
\]
and therefore emphasizes the compression/expansion structure associated with the internal shock-cell-like response. In contrast, $Kn_{\rm GLL}$ is an unsigned composite maximum over several fields. It combines wall-normal thermal gradients, pressure and density stratification, Mach-number variation, throat acceleration, and the downstream compression structure into a single positive indicator. Under cooling and adiabatic conditions, the compression feature remains relatively compact and can still appear as a localized short-gradient-length region. Under heating, the wall-driven thermal and viscous layers become strong enough that they dominate the composite indicator over much of the diverging wall. The shock-cell-like compression is therefore not absent; it is no longer the only or dominant short-gradient-length feature in the composite map.

The evolution from cooling to heating in Fig.~\ref{fig:kngll} is consequently not simply a weakening of the shock. It is a redistribution of local non-equilibrium. Cooling preserves a wider low-gradient core and a more localized compression-related contribution. Heating progressively expands the wall-dominated high-$Kn_{\rm GLL}$ region, narrows the low-gradient central core, and spreads the compression adjustment into a broader viscous--thermal environment. This trend supports the main mechanism of the paper: prescribed wall heating simultaneously modifies the thermal field, the local transport scale, and the effective aerodynamic passage. The result is a coupled rarefaction--blockage response, rather than a purely inviscid shock displacement.

\subsection{Effective aerodynamic blockage}

\begin{figure}[H]
\centering
\includegraphics[width=0.96\textwidth]{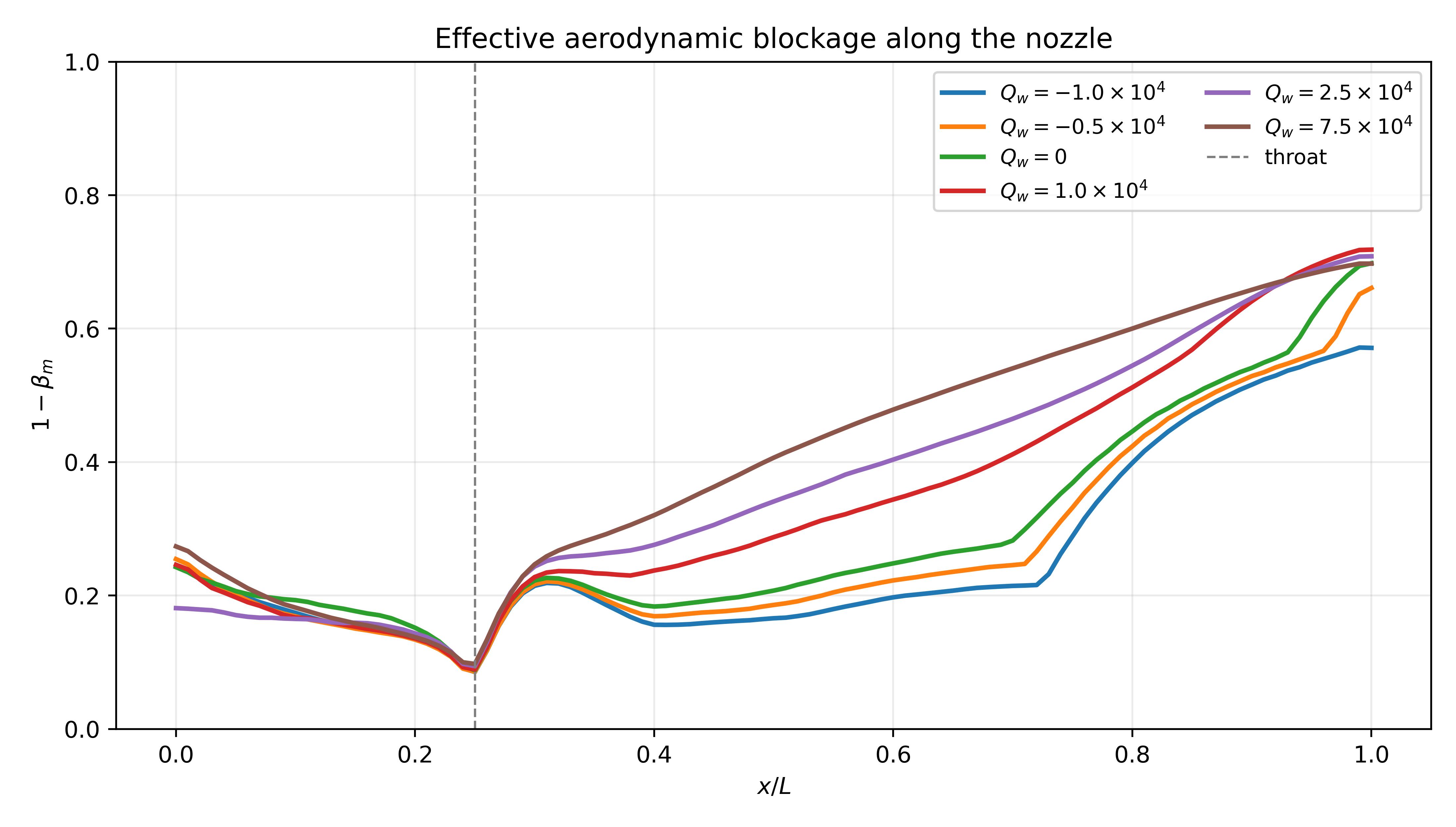}
\caption{Effective aerodynamic blockage, $1-\beta_m$, along the nozzle. Heating produces a progressive increase in blockage throughout the diverging section, demonstrating contraction of the effective mass-carrying core.}
\label{fig:blockage_profiles}
\end{figure}

The near-wall slip response leads naturally to the blockage analysis. The strongest new diagnostic result is shown in Fig.~\ref{fig:blockage_profiles}. The effective blockage, $1-\beta_m$, is small near the throat for all cases but grows downstream in the diverging section. Cooling maintains a relatively wide mass-carrying core, while heating progressively increases blockage. In the strongest heating case, the blockage exceeds 0.5 near the exit. This means that less than half of the local half-height is effectively carrying mass flux at the level implied by the peak axial mass-flux density.

The blockage profile explains why heating reduces the discharge coefficient even though it adds energy to the gas. The heat input raises the wall temperature and, as shown by Fig.~\ref{fig:slip}, modifies the near-wall tangential momentum response. The resulting viscous--thermal layer occupies a larger part of the diverging section and shifts the axial mass flux toward a narrower core. The nozzle therefore behaves as if its aerodynamic passage were smaller than its geometric passage. This effect is familiar in microflows as a strong surface-to-volume consequence, but the present definition measures it directly from DSMC fields.

\begin{figure}[H]
\centering
\includegraphics[width=0.96\textwidth]{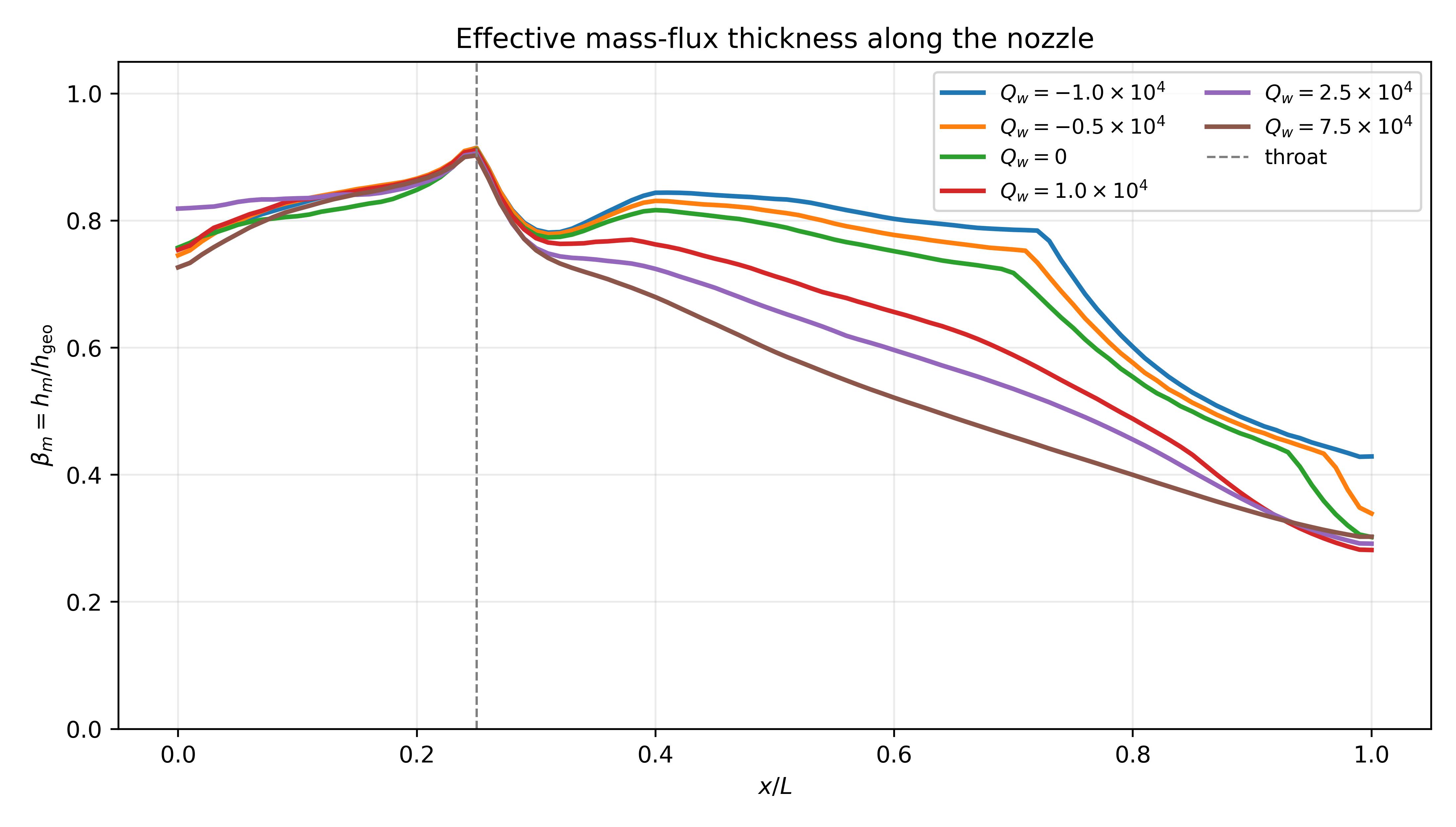}
\caption{Effective mass-flux thickness, $\beta_m=h_m/h_{\rm geo}$, along the nozzle. The reduction of $\beta_m$ downstream of the throat under heating provides direct evidence that wall heat flux contracts the mass-carrying core.}
\label{fig:thickness_profiles}
\end{figure}

The same mechanism is shown from the complementary perspective of the mass-flux thickness in Fig.~\ref{fig:thickness_profiles}. Downstream of the throat, $\beta_m$ decreases strongly under heating. The cooling cases preserve $\beta_m$ at higher values over most of the diverging section, whereas strong heating causes a monotonic contraction toward the outlet. The decrease of $\beta_m$ is the field-level origin of the reduced mass flow rate. It is not simply a post-processing correlation with $C_d$; it is computed directly from the distribution of $\rho u$ across the nozzle height.

\subsection{Thermal--aerodynamic trade-off}

\begin{figure}[H]
\centering
\includegraphics[width=0.96\textwidth]{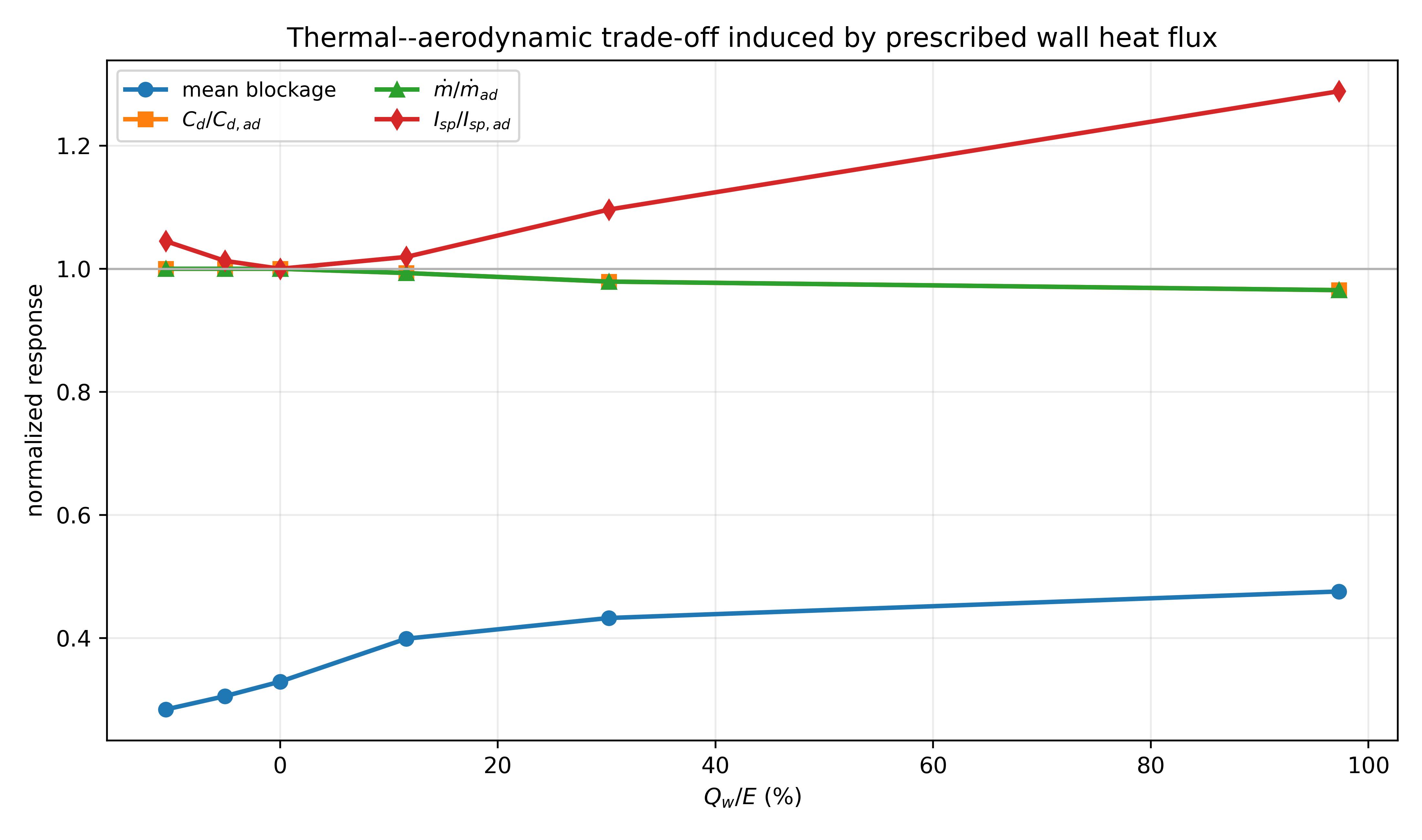}
\vspace{0.6em}

\includegraphics[width=0.96\textwidth]{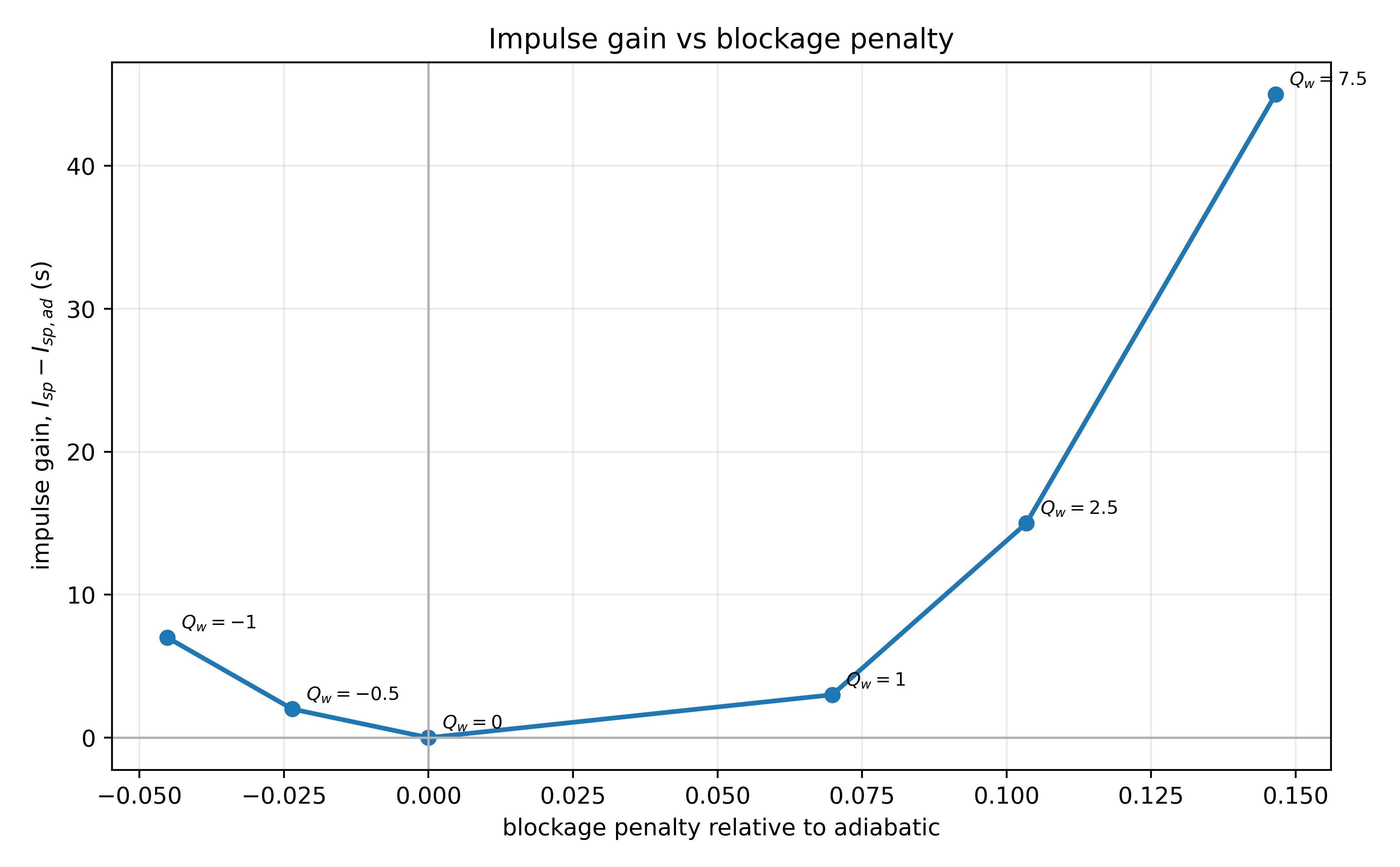}
\caption{Thermal--aerodynamic trade-off induced by prescribed wall heat flux. (a) Variation of mean effective blockage, normalized discharge coefficient, normalized mass-flow rate, and normalized specific impulse with $Q_w/E$. Heating increases blockage and reduces throughput, but the impulse can rise because thermal augmentation changes the thrust per unit mass flow. (b) Impulse gain plotted against blockage penalty. Strong heating moves toward larger blockage penalty and larger impulse gain, showing that thermal augmentation can outpace the mass-flow penalty when the objective is impulse per unit mass.}
\label{fig:tradeoff}
\end{figure}

The blockage mechanism is connected to the integrated performance metrics in Fig.~\ref{fig:tradeoff} and Table~\ref{tab:performance}. Figure~\ref{fig:tradeoff}(a) uses $Q_w/E$ as the horizontal coordinate, so the thermal forcing is measured relative to the inlet kinetic-energy flux rather than by the dimensional heat flux alone. The figure shows the coupled aerodynamic penalty and impulse benefit. As heating increases, the mean effective blockage rises, the discharge coefficient decreases, and the normalized mass-flow rate drops. At the same time, sufficiently strong heating increases the normalized specific impulse. The strongest heating case produces $I_{sp}=201~{\rm s}$, compared with $156~{\rm s}$ in the adiabatic case, even though the mass flow rate decreases from $1.44\times10^{-7}$ to $1.39\times10^{-7}~{\rm kg\,s^{-1}}$.

Figure~\ref{fig:tradeoff}(b) presents the same trade-off in a more diagnostic form by plotting impulse gain against blockage penalty. This panel is useful because it separates the two competing effects that are otherwise mixed in a single heat-flux sweep. Points to the right have a larger blockage penalty, while points upward have a larger impulse gain. The strong-heating case moves into the upper-right part of the map: it pays an aerodynamic cost through mass-carrying-core contraction, but gains impulse because the thermal and pressure-thrust contributions increase more than the mass-flow penalty. The cooling and weak-heating cases remain closer to the adiabatic reference because their thermal forcing is too small to reorganize the thrust-per-unit-mass balance strongly.

This is the main physical result of the paper. Wall heat flux produces a trade-off rather than a one-sided performance change. The aerodynamic penalty is the contraction of the effective passage and the reduction of mass throughput. The thermodynamic benefit is enthalpy augmentation and pressure-thrust increase. The underlying thrust decomposition, mass-flow rate, and specific impulse values used in Fig.~\ref{fig:tradeoff} are listed in Table~\ref{tab:performance}. Therefore, the design implication depends on the objective. If maximum mass throughput is desired, strong heating is unfavorable. If high specific impulse is desired and a moderate loss of mass flow is acceptable, wall heating can be beneficial. The combined trade-off map is therefore more informative than separate plots of $C_d$, $\dot{m}$, and $I_{sp}$.

\begin{table}[t]
\centering
\caption{Propulsive performance metrics extracted from the DSMC solutions. Thrust components are normalized by $P_iA_t$, and mass flux is reported in $10^{-7}~{\rm kg\,s^{-1}}$.}
\label{tab:performance}
\begin{tabular}{lrrrrrrr}
\toprule
Case & $Q_w$ & $Q_w/E$ & $F_{\rm mom}$ & $F_{\rm press}$ & $F_t$ & $\dot{m}\times10^7$ & $I_{sp}$ \\
 & $(10^4{\rm W\,m^{-2}})$ & (\%) & $/P_iA_t$ & $/P_iA_t$ & $/P_iA_t$ & & (s) \\
\midrule
Cooling 1 & -1.0 & -10.5 & 1.60 & 1.66 & 3.26 & 1.44 & 163 \\
Cooling 2 & -0.5 & -5.06 & 1.45 & 1.69 & 3.14 & 1.44 & 158 \\
Adiabatic & 0 & 0 & 1.30 & 1.81 & 3.11 & 1.44 & 156 \\
Heating 1 & 1.0 & 11.6 & 1.20 & 1.95 & 3.15 & 1.43 & 159 \\
Heating 2 & 2.5 & 30.2 & 1.24 & 2.11 & 3.35 & 1.41 & 171 \\
Heating 3 & 7.5 & 97.3 & 1.58 & 2.30 & 3.89 & 1.39 & 201 \\
\bottomrule
\end{tabular}
\end{table}

\subsection{Wall--bulk heat-transfer scaling: Nusselt and Brinkman responses}

\begin{figure}[!htbp]
\centering
\begin{subfigure}{0.98\textwidth}
\centering
\includegraphics[width=\textwidth]{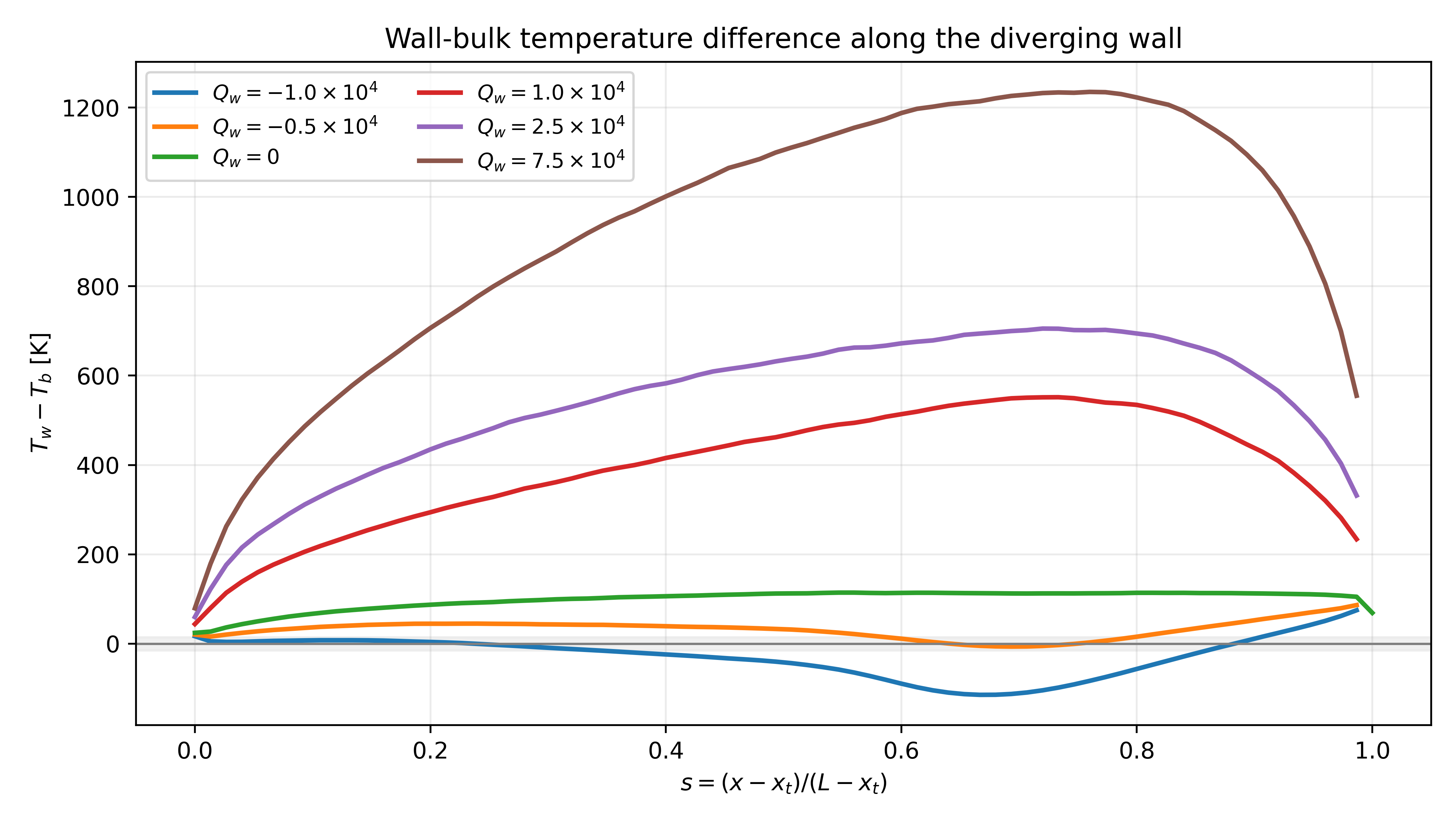}
\caption{Wall--bulk temperature difference and the singular-denominator band.}
\label{fig:deltat}
\end{subfigure}
\vspace{0.6em}
\begin{subfigure}{0.98\textwidth}
\centering
\includegraphics[width=\textwidth]{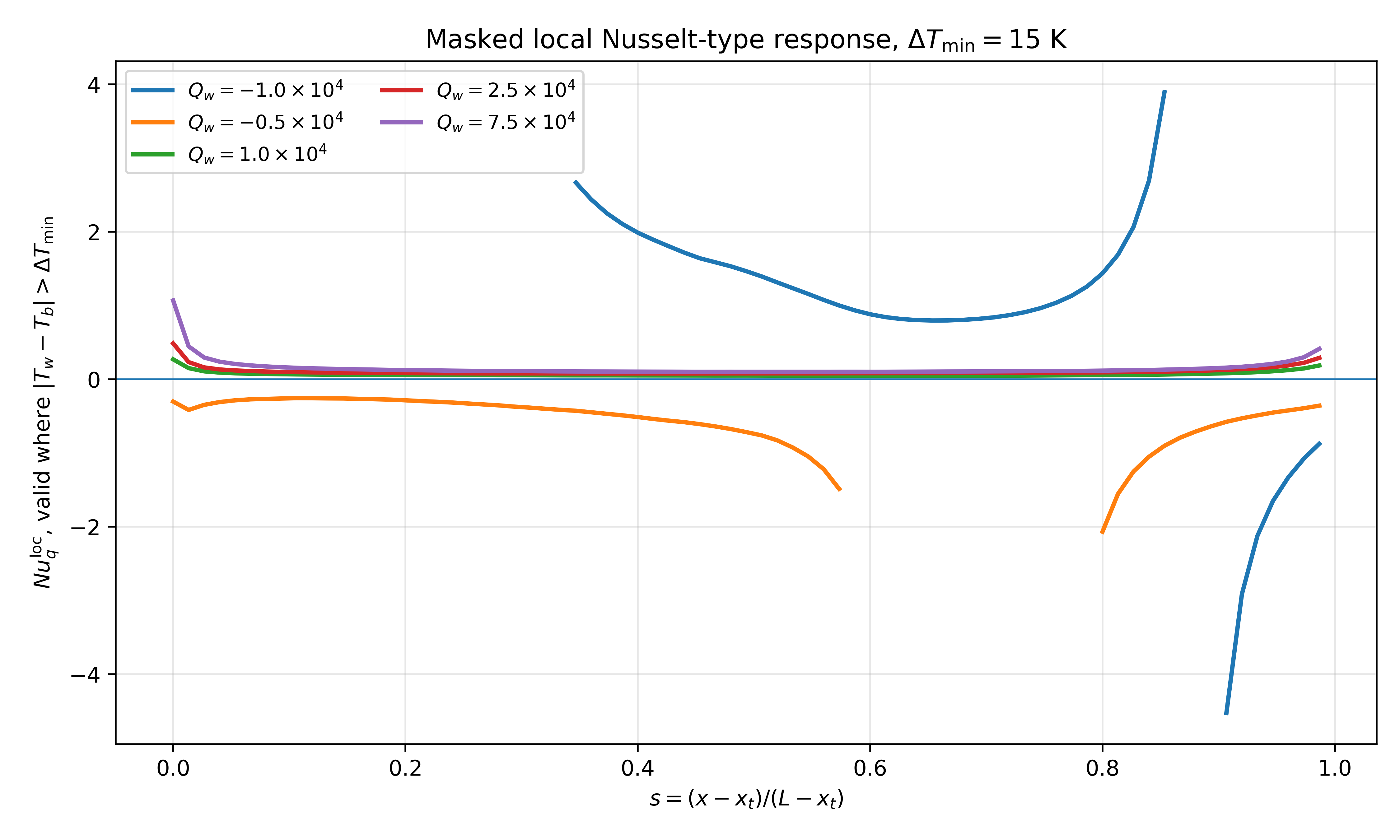}
\caption{Finite signed local Nusselt-type response after applying only the validity mask.}
\label{fig:nuvalid}
\end{subfigure}
\caption{Wall--bulk thermal conditioning and local Nusselt-type response on Wall-2. Panel (a) identifies the wall--bulk temperature crossings where an imposed-flux-to-temperature-difference ratio becomes singular. The shaded band denotes $|T_w-T_b|\le0.05T_0$ and is used only as a validity mask. Panel (b) reports the signed local Nusselt-type response using the VHS/Wu--Eucken temperature-dependent conductivity. The mask is a validity mask for the ratio definition, not a smoothing or denoising operation.}
\label{fig:heat_transfer_diag}
\end{figure}

The Nusselt-type diagnostic is controlled by two quantities: the wall--bulk temperature difference and the local film-temperature conductivity used in Eq.~\eqref{eq:nuq}. The wall--bulk temperature difference determines whether the imposed-flux-to-temperature-difference ratio is well conditioned, while the conductivity provides the temperature-dependent transport scaling. The conductivity itself is not plotted as a separate main-text figure because its role is auxiliary: it is already specified by Eq.~\eqref{eq:nuq} through the VHS/Wu--Eucken relation and follows directly from the film temperature. Instead, Fig.~\ref{fig:heat_transfer_diag} combines the two quantities that determine the interpretability of the Nusselt response: panel (a) shows the denominator behavior through $T_w-T_b$, and panel (b) shows the resulting finite signed local Nusselt-type response after applying only the singular-denominator validity mask.

Figure~\ref{fig:deltat} shows $T_w-T_b$ along the thermally forced diverging wall, where $s=(x-x_t)/(L-x_t)$ measures the normalized distance from the throat to the exit. Heating produces a positive wall--bulk separation over almost the entire diverging wall. The strongest heating case reaches $T_w-T_b\simeq 1200~{\rm K}$ over the downstream portion of Wall-2, whereas the moderate heating cases remain substantially lower. The adiabatic case has a finite positive wall--bulk offset because the expanding gas cools relative to the wall even when no net wall heat flux is prescribed.

Cooling is qualitatively different. For $Q_w=-1\times10^4~{\rm W\,m^{-2}}$, $T_w-T_b$ crosses zero twice in the diverging section, and for $Q_w=-0.5\times10^4~{\rm W\,m^{-2}}$ the crossing occurs in the downstream part of Wall-2. These zero crossings are not numerical details; they mark locations where any imposed-flux-to-temperature-difference Nusselt ratio becomes ill-conditioned. The shaded band in Fig.~\ref{fig:deltat} identifies the $|T_w-T_b|\le 0.05T_0$ region that is excluded only from the comparative finite-Nusselt plot in Fig.~\ref{fig:nuvalid}. The raw signed response, including the singular spikes, is retained in ~\ref{app:raw_nusselt}.

The signed local Nusselt-type response in Fig.~\ref{fig:nuvalid} is evaluated with the temperature-dependent conductivity from Eq.~\eqref{eq:nuq}. The conductivity is computed at the wall--bulk film temperature using $f_{{\rm eu},N_2}=1.96$ and $f_{\rm dof}=5$, i.e. $k_f=4.90(k_B/m)\mu_f$. This property variation is important because the strongest heating cases change the film temperature substantially: cooling lowers $k_f$ over much of Wall-2, whereas strong heating increases it downstream. However, because this conductivity trend mirrors the already-shown wall--bulk thermal response, it is not given a separate figure. The key heat-transfer result is the conditioned ratio in Fig.~\ref{fig:nuvalid}. The cooling cases remain piecewise and sign-sensitive because their wall--bulk temperature difference changes sign. Heating cases are much better conditioned: $Nu_q^{\rm loc}$ stays positive and low-amplitude because the imposed heat flux is accompanied by a large wall--bulk temperature difference and an increased film conductivity. Thus strong heating does not imply proportionally stronger wall-to-core thermal coupling; much of the energy is absorbed into a hot near-wall layer while the mass-carrying core responds more weakly.

\begin{figure}[!htbp]
\centering
\begin{subfigure}{0.98\textwidth}
\centering
\includegraphics[width=\textwidth]{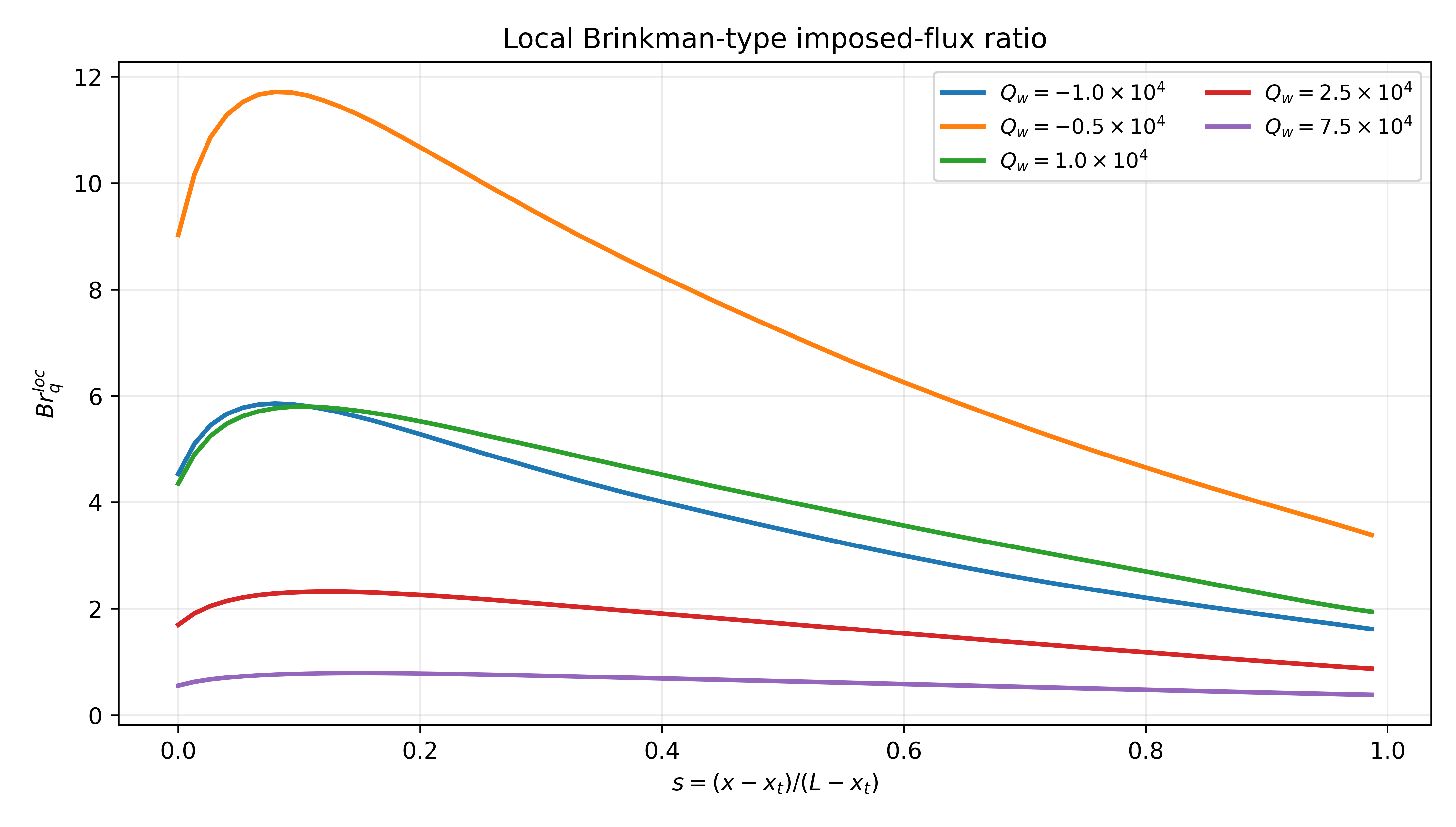}
\caption{Local-viscosity Brinkman-type imposed-flux ratio.}
\label{fig:brprofiles}
\end{subfigure}

\vspace{0.6em}

\begin{subfigure}{0.98\textwidth}
\centering
\includegraphics[width=\textwidth]{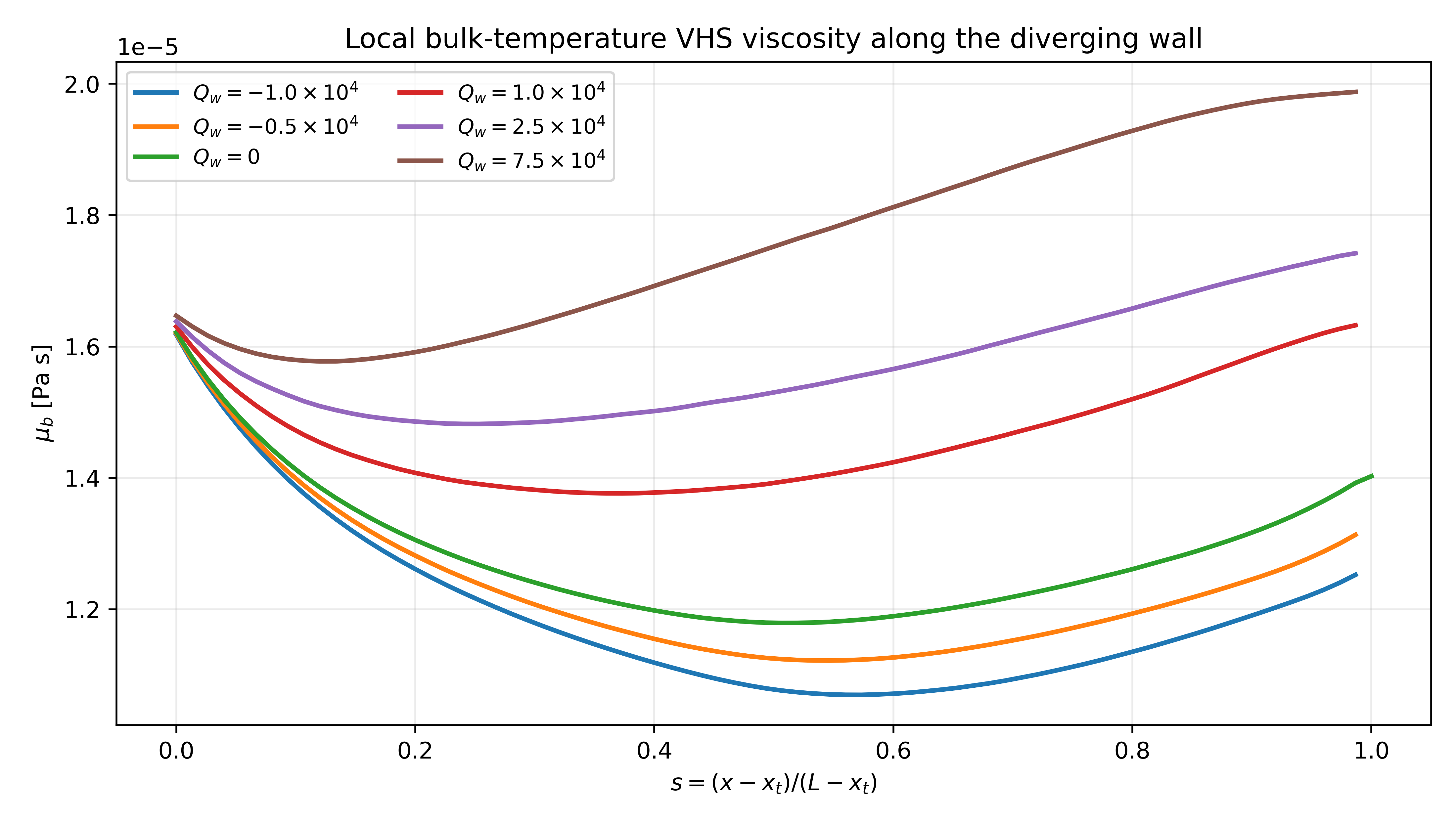}
\caption{Bulk-temperature-based VHS viscosity used in the Brinkman numerator.}
\label{fig:mulocal}
\end{subfigure}

\caption{Coupled viscous--thermal scaling on the thermally forced diverging wall. Panel (a) reports the local Brinkman-type imposed-flux ratio, $Br_q^{\rm loc}=\mu_bU_b^2/(|Q_w|D_h+\epsilon_Q)$, comparing the local viscous momentum-transport scale with the imposed wall-thermal scale. Panel (b) shows the corresponding bulk-temperature-based VHS viscosity, $\mu_b=\mu_{\rm ref}(T_b/T_{\rm ref})^\omega$, used in panel (a). The decrease of $Br_q^{\rm loc}$ under strong heating is therefore not caused by using a fixed viscosity; it persists even when the local viscosity increase produced by wall heating is included. Because this diagnostic scales the local viscous momentum-transport scale by the imposed heat-flux magnitude, the adiabatic case, $Q_w=0$, is not plotted or interpreted through $Br_q^{\rm loc}$; it is used only as the thermal baseline in the other diagnostics.}
\label{fig:br_mu_combined}
\end{figure}

The coupled Brinkman--viscosity diagnostic in Fig.~\ref{fig:br_mu_combined} provides the complementary momentum--thermal interpretation. Panel~\ref{fig:brprofiles} shows the local-viscosity Brinkman-type imposed-flux ratio,
\begin{equation}
Br_q^{\rm loc}(x)=
\frac{\mu_b(x)U_b^2(x)}
{|Q_w|D_h(x)+\epsilon_Q},
\label{eq:brq_discussion}
\end{equation}
whereas panel~\ref{fig:mulocal} shows the corresponding bulk-temperature-based viscosity used in the numerator,
\begin{equation}
\mu_b(x)=\mu_{\rm ref}\left(\frac{T_b(x)}{T_{\rm ref}}\right)^\omega .
\label{eq:mub_discussion}
\end{equation}
The two panels should be interpreted together. The viscosity panel confirms that transport-property variation is retained: cooling lowers the bulk temperature and therefore decreases $\mu_b$, while heating increases $\mu_b$ downstream as wall energy penetrates into the gas. The Brinkman panel then shows that this viscosity increase is not sufficient to offset the much larger imposed thermal scale $|Q_w|D_h$ in the denominator.

This structure explains the non-monotonic ordering of the curves in Fig.~\ref{fig:br_mu_combined}. The weak-cooling case, $Q_w=-0.5\times10^4~{\rm W\,m^{-2}}$, gives the largest $Br_q^{\rm loc}$ because $|Q_w|$ is small, so the denominator is small relative to the local viscous momentum scale. The stronger cooling case, $Q_w=-1.0\times10^4~{\rm W\,m^{-2}}$, has a larger denominator and a colder, lower-viscosity bulk gas, so its ratio is lower despite having the same sign of heat transfer. Weak heating can remain comparable to cooling in the upstream part of the diverging section because the heat-flux magnitude is still modest and the local velocity scale remains large.

As heating is increased, the rise in $\mu_b$ is not sufficient to compensate for the much larger imposed thermal scale in $|Q_w|D_h$. Consequently, $Br_q^{\rm loc}$ decreases from weak heating to strong heating even though the gas viscosity itself increases. The strongest heating case, $Q_w=7.5\times10^4~{\rm W\,m^{-2}}$, remains below unity over most of Wall-2, indicating that the imposed wall heat flux is larger than the local viscous momentum-transport scale. Thus the combined figure demonstrates that the strongest heating case enters a thermally dominated regime even after the local VHS viscosity increase is included. This regime is the same one in which the wall--bulk temperature separation is largest, the effective mass-flux thickness contracts, and the specific impulse increases.

Together, Figs.~\ref{fig:walltemp}, \ref{fig:heat_transfer_diag}, and \ref{fig:br_mu_combined} provide the heat-transfer mechanism behind the aerodynamic trends. Wall heating raises $T_w$, but the mass-flux-weighted bulk temperature responds more slowly because the near-wall viscous--thermal layer thickens and shields the core. The local Nusselt-type response decreases for strong heating even as the imposed heat flux increases, because both the wall--bulk temperature difference and the film-temperature-based conductivity increase. At the same time, the local-viscosity Brinkman-type ratio decreases with heating, indicating that the imposed wall heat flux becomes dynamically dominant relative to the local viscous momentum scale. Thus $Nu_q^{\rm loc}$, $Br_q^{\rm loc}$, $k_f$, $\mu_b$, $\beta_m$, and $I_{sp}$ are different projections of the same rarefied viscous--thermal mechanism: heating creates a hot, low-momentum near-wall layer that narrows the effective mass-carrying passage while adding enough thermal and pressure thrust to increase specific impulse at high heat input.

\subsection{Low-dimensional organization of the thermal response}

POD is applied to the signed numerical-schlieren fields in order to determine whether the heat-flux-parametric response is a random DSMC fluctuation pattern or a coherent structural deformation of the rarefied nozzle flow. For each heat-flux case,
\[
Q_w=\{-1.0,-0.5,0,1.0,2.5,7.5\}\times 10^4~{\rm W\,m^{-2}},
\]
one statistically converged DSMC density field is first interpolated onto the same post-processing grid. The signed numerical-schlieren field is then computed as
\[
\chi(x,y;Q_w)=
\frac{\partial(\rho/\rho_0)}{\partial(x/L)} ,
\]
so that positive and negative density-gradient structures are retained instead of being collapsed into an unsigned magnitude. Each snapshot used in the POD is therefore a full two-dimensional signed schlieren field, not a one-dimensional wall profile or an integrated performance quantity. The six snapshots correspond to the cooling, adiabatic and heating members of the same nozzle family. Before the decomposition, the ensemble mean field is subtracted,
\[
\chi'(x,y;Q_w)=\chi(x,y;Q_w)-\overline{\chi}(x,y),
\]
and each fluctuation field is vectorized to form the snapshot matrix
\[
\mathbf{X}=\left[
\chi'_1,\chi'_2,\ldots,\chi'_6
\right].
\]
The POD modes are obtained from the singular value decomposition
\[
\mathbf{X}=\mathbf{U}\boldsymbol{\Sigma}\mathbf{V}^{T},
\]
where the columns of $\mathbf{U}$ define the spatial modes $\phi_k$, the singular values determine the modal energies, and the columns of $\mathbf{V}$ give the heat-flux-dependent modal coefficients. The relative energy of mode $k$ is evaluated as
\[
E_k=\frac{\sigma_k^2}{\sum_j \sigma_j^2}.
\]
This procedure means that the POD basis is global with respect to the heat-flux sweep: the cooling, adiabatic and heating cases are not decomposed separately, and no case-specific shock location or hand-tuned registration is imposed.

\begin{figure}[H]
\centering
\includegraphics[width=0.92\textwidth]{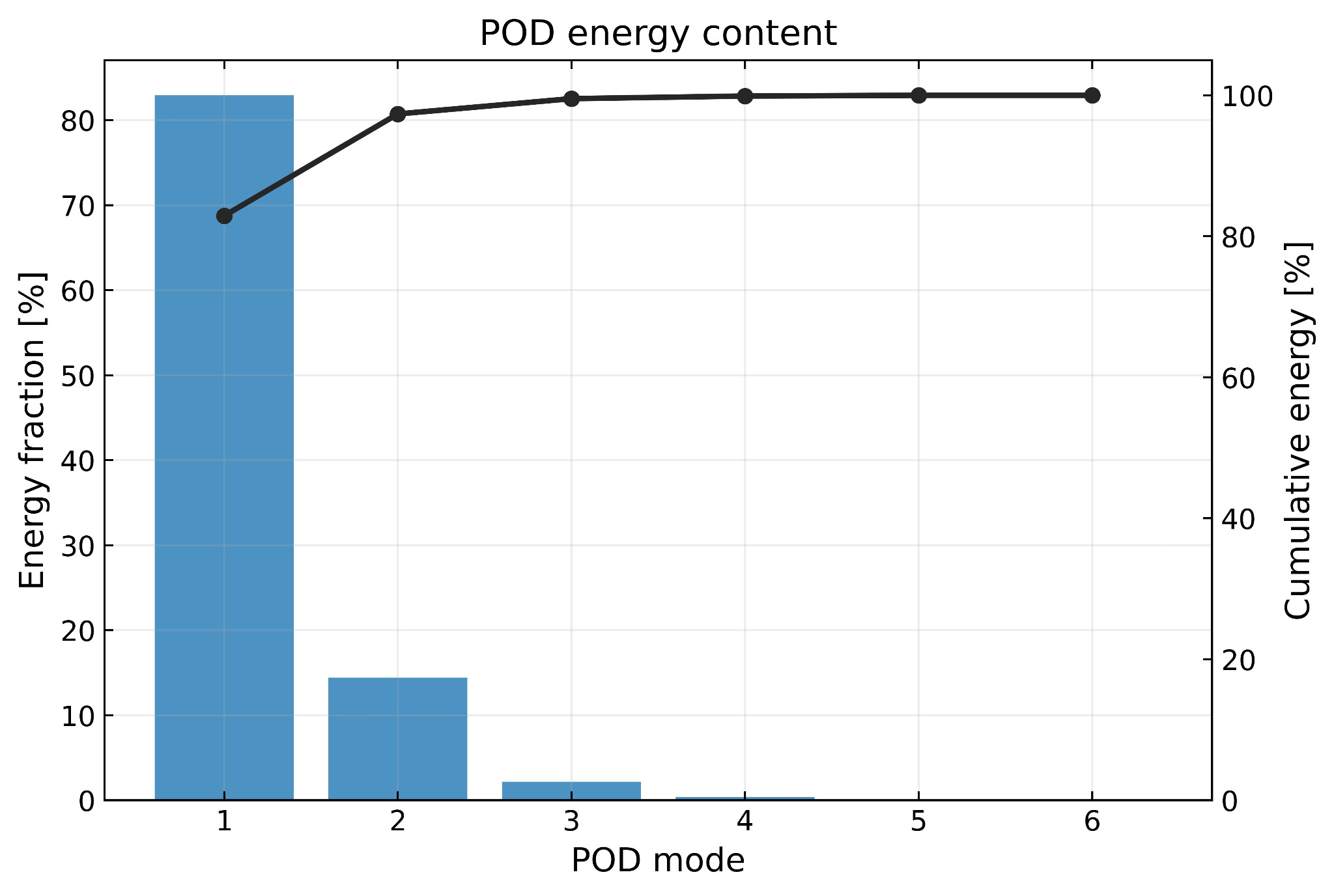}
\caption{POD energy spectrum of the signed numerical-schlieren field. The first mode contains 82.91\% of the fluctuation energy, while the first two modes contain more than 97\%.}
\label{fig:pod_energy}
\end{figure}

The energy spectrum in Fig.~\ref{fig:pod_energy} shows that the first mode contains 82.91\% of the fluctuation energy and the first two modes contain more than 97\%. The first four modes essentially capture the full family. This compactness is not as extreme as in the earlier specular-wall, shock-centered micro-nozzle problem, where registration and jump scaling made the internal compression layer nearly one-mode~\citep{roohi2025shockfusion}. The difference is physically meaningful. In the earlier problem, the wall was specular and the leading parametric variation was mainly the displacement and finite thickness of an internal compression layer. In the present problem, the diffuse wall and the prescribed wall heat flux introduce additional coherent degrees of freedom: wall-layer thickening, thermal shielding of the core, compression weakening and outlet-region reorganization. Four modes are therefore not a weakness of the analysis; they show that the flow remains low-dimensional but is richer than a single moving compression feature.

\begin{figure}[H]
\centering
\includegraphics[width=1.1\textwidth]{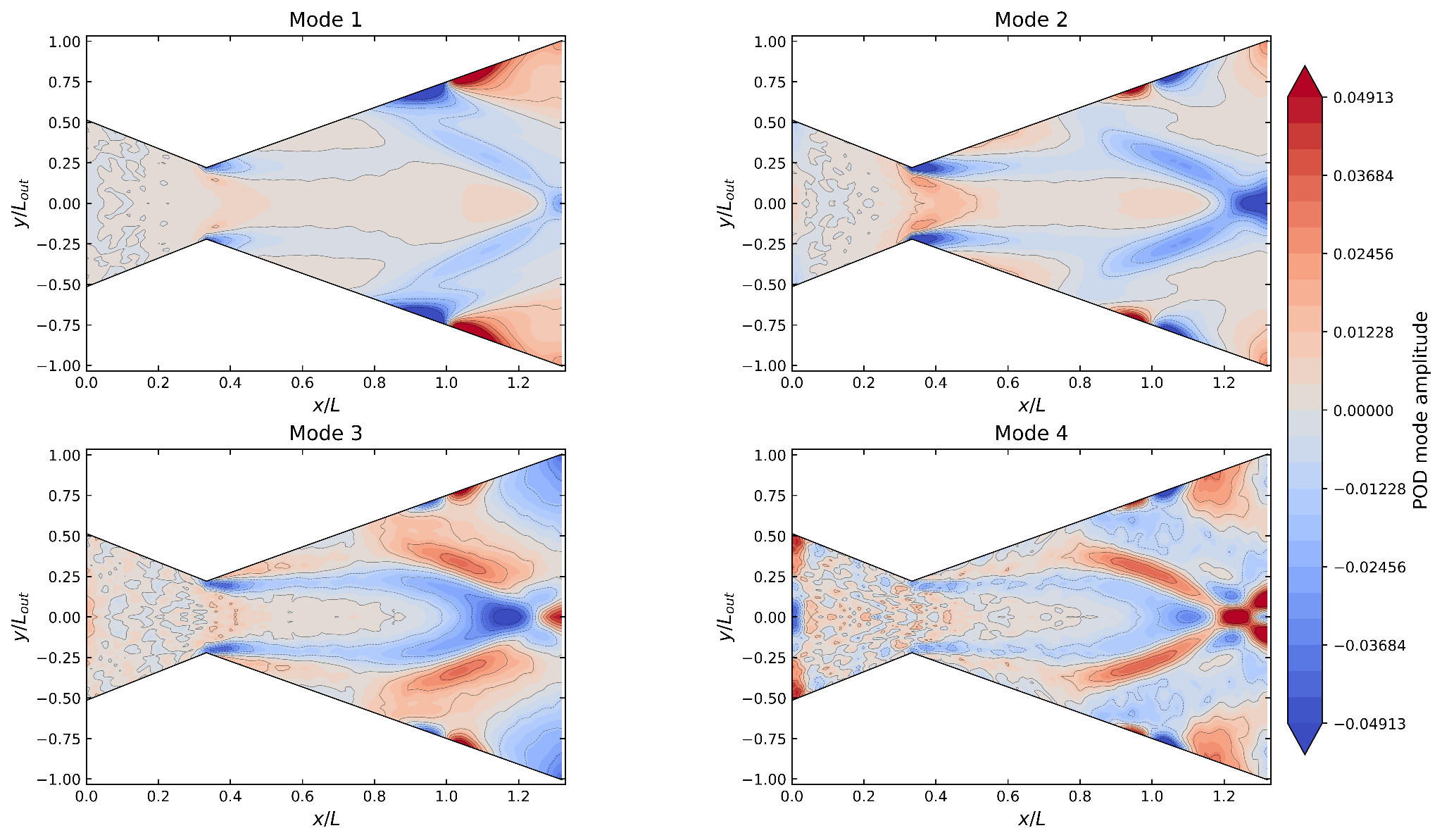}
\caption{First four POD modes of the signed numerical-schlieren field. The modes are concentrated near the throat, diverging wall and outlet compression region. The leading modes describe the weakening and reorganization of the compression signature, while the higher modes carry smaller-scale corrections associated with the wall-adjacent viscous--thermal layer and the interface between the fast core and the heated near-wall flow.}
\label{fig:pod_modes}
\end{figure}

The spatial modes in Fig.~\ref{fig:pod_modes} provide the structural interpretation of this compactness. Mode 1 represents the dominant weakening and redistribution of the density-gradient field as wall heating transforms the relatively compact internal compression feature into a broader viscous--thermal compression zone. Mode 2 contains a complementary outlet-region structure, indicating that the downstream compression-cell response is not a passive extension of the leading mode but a coherent part of the heat-flux-induced reorganization.

Modes 3 and 4 have smaller energy, but their spatial support is physically important. They are concentrated near the diverging wall, the throat-to-diverging transition and the residual outlet compression pattern. These modes should not be interpreted as independent flow phenomena in a strict one-mode/one-physics sense, because POD modes are orthogonal mathematical structures and their signs are arbitrary. Nevertheless, their location suggests a plausible physical role: they carry the smaller-scale correction associated with the shear/transition layer between the fast, relatively colder core flow and the slower, strongly heated wall-adjacent layer. In this interpretation, modes 3 and 4 capture the adjustment of the thermal-stratification interface rather than only numerical residue. The fact that these modes are localized where the wall-driven mass-flux layer, $Kn_{\rm GLL}$ enhancement and signed density-gradient redistribution overlap indicates that POD is detecting the physical boundaries of the thermally layered flow.

\begin{figure}[H]
\centering
\includegraphics[width=1.1\textwidth]{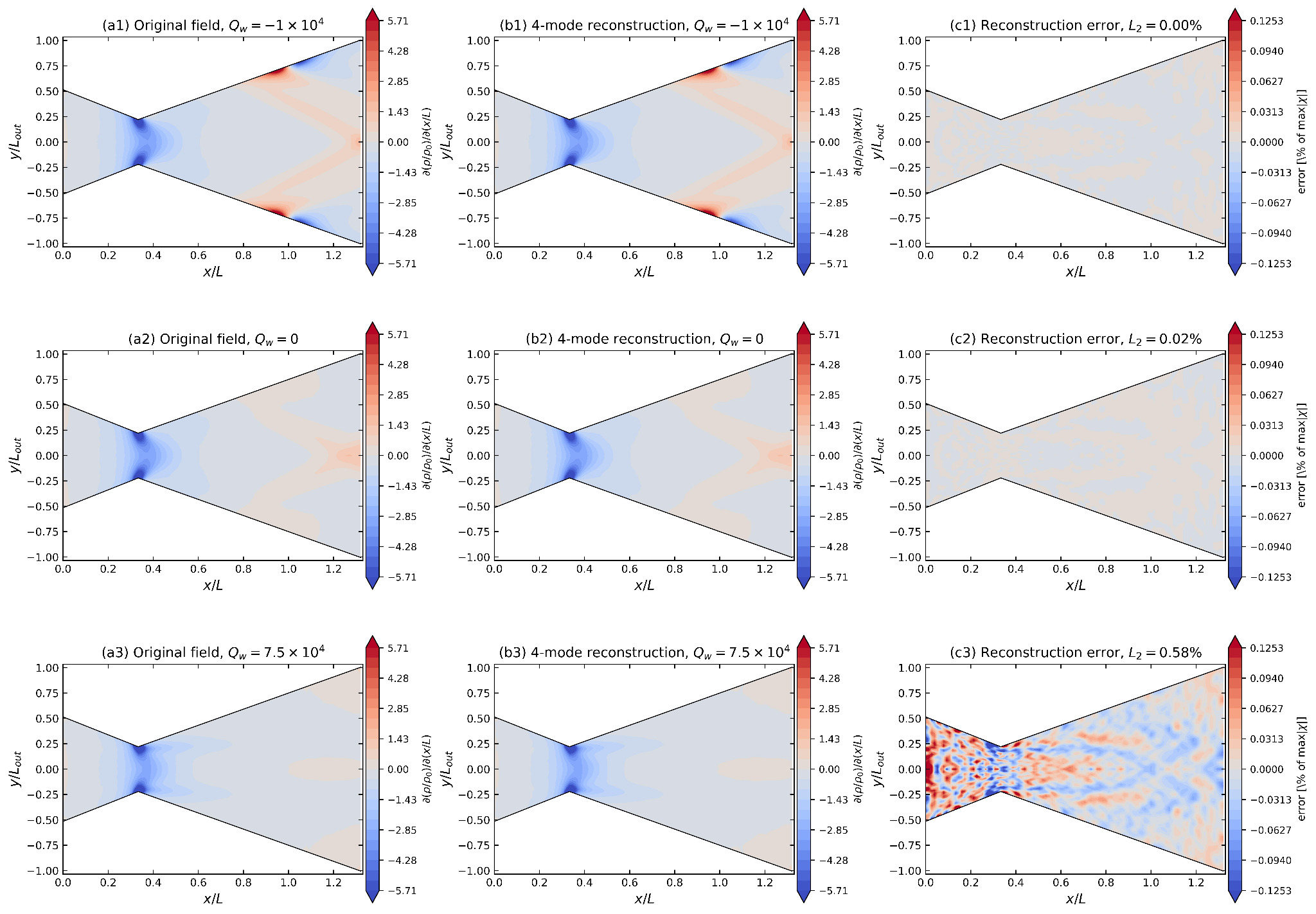}
\caption{Four-mode POD reconstruction of representative cooling, adiabatic and strong-heating numerical-schlieren fields. The sub-percent reconstruction errors show that the full heat-flux family is organized by a small number of coherent modes, even though the thermally forced wall layer requires more modal content than a purely shock-displacement-dominated specular-wall nozzle.}
\label{fig:pod_recon}
\end{figure}

The four-mode reconstruction in Fig.~\ref{fig:pod_recon} demonstrates the practical meaning of the POD spectrum. Representative cooling, adiabatic and strong-heating fields are reconstructed with sub-percent relative errors. This result is important because it shows that the low-rank property is not confined to one selected case. The global POD basis, built from all six heat-flux snapshots, represents the full family with only a few modes. The adiabatic and strong-heating cases are not reconstructed by imposing a shock location or by using separate case-specific bases; they are reconstructed by the same heat-flux-family basis.

This finding should be read in the same spirit as recent rarefied shock-layer compactness analyses. In the bow-shock problem, density can become nearly rank one after density-attached registration, whereas Mach and thermal variables retain independent modal content. In the present nozzle, the signed schlieren field is low-dimensional but not one-dimensional. The difference is a physical consequence of the wall-heat-flux boundary condition: thermal forcing changes both the compression signature and the wall-induced mass-flux layer. Thus, POD supports the main mechanism of the paper rather than replacing it. It shows that the heat-flux response is not random DSMC scatter and not a collection of unrelated fields, but a coherent deformation of the compression structure and the wall-driven viscous--thermal layer.

\subsection{Implications for thermal control of rarefied micro-nozzles}

The combined diagnostics support a consistent physical picture. Wall heating first changes the energy of reflected molecules. This raises the wall temperature and modifies near-wall molecular transport. The near-wall low-momentum layer expands into the diverging section, reducing $\beta_m$ and increasing $1-\beta_m$. This produces a smaller effective aerodynamic passage, explaining the decline in mass flux and discharge coefficient. At the same time, the gas receives additional enthalpy, and the pressure contribution to thrust increases. At high heat input, the impulse gain exceeds the throughput penalty.

This means that the best operating point depends on the mission metric. For mass-flow-limited applications, strong wall heating may be undesirable. For specific impulse or energy-conversion-focused applications, strong heating may be beneficial. The present results therefore do not support a universal statement that wall heating improves micro-nozzle performance. They support a more precise statement: wall heating trades mass-carrying area for thermal impulse gain. The value of this trade-off depends on available wall power, allowable thermal load, material limits and required mass throughput.

\subsection{Detailed mechanism of mass-flux-thickness contraction}

The blockage metric can be understood by comparing two limiting mass-flux profiles at the same geometric half-height. If the axial mass flux is nearly uniform across the half-channel, then $\int \rho u\,dy\approx h_{\rm geo}(\rho u)_{\max}$ and $\beta_m\approx 1$. If the axial flux is concentrated in a narrow core because the near-wall region carries little axial momentum, then the same integral becomes much smaller than $h_{\rm geo}(\rho u)_{\max}$ and $\beta_m$ decreases. Thus, $\beta_m$ is a mass-flux-weighted analog of an effective flow width. It does not require an arbitrary edge definition and does not require a continuum boundary-layer profile.

This distinction matters in DSMC data. The near-wall region is not a classical no-slip boundary layer with a smooth continuum velocity profile. The wall-reflected molecular distribution is affected by the wall temperature, by the local accommodation model, by the local collision frequency, and by particle sampling noise. A thickness based on a velocity threshold could therefore be sensitive to the chosen threshold. The integral definition in Eq.~\eqref{eq:hm} is more robust: it measures how much cross-sectional width is needed to carry the observed mass flux at the local maximum flux density. The same metric also connects directly to the global mass-flow rate because integrating $j_m=\rho u$ over the section is the local mass flow per unit depth.

The combined behavior in Figs.~\ref{fig:slip}, \ref{fig:blockage_profiles} and \ref{fig:thickness_profiles} also reveals why the discharge coefficient decreases even though the strongest heating case receives substantial thermal energy. Heating raises the wall-adjacent molecular energy and increases thermal diffusion. It also reduces density near the hot wall and changes the local momentum distribution. The combined result is a wall layer that occupies a larger fraction of the diverging passage. In a macroscopic nozzle, an energy addition might be interpreted mainly through stagnation-enthalpy increase. In the present micro-nozzle, the same energy addition also changes the effective flow area. This is the central micro-scale difference: because the surface-to-volume ratio is large, thermal control is inevitably also geometric control in an aerodynamic sense.

The non-monotonic intermediate behavior around $Q_w=1\times10^4~{\rm W\,m^{-2}}$ and $2.5\times10^4~{\rm W\,m^{-2}}$ should not be overinterpreted as numerical scatter. The field response is not controlled by one scalar parameter alone. Heat flux modifies the wall temperature, the local viscosity, the density level, the compression pattern, and the exit pressure distribution simultaneously. A moderate heating case can increase blockage locally while also shifting the region of maximum mass flux. The robust trend is the contrast between cooling/adiabatic conditions and strong heating: strong heating produces a larger wall-affected low-momentum region and a smaller effective mass-carrying core.

\subsection{Why the specific impulse increases despite blockage}

The apparent contradiction between lower mass flux and higher specific impulse is resolved by separating throughput from impulse efficiency. The mass flow rate is determined by the integrated axial mass flux,
\begin{equation}
\dot{m}=\int_A \rho u_x\,dA.
\end{equation}
It is reduced by heating because the effective mass-carrying width decreases. The specific impulse, however, is
\begin{equation}
I_{sp}=\frac{F_{\rm mom}+F_{\rm press}}{\dot{m}g_0},
\end{equation}
where
\begin{equation}
F_{\rm mom}=\int_A\rho u_x^2\,dA,\qquad
F_{\rm press}=\int_A(p-p_{\rm amb})\,dA.
\end{equation}
Thus, $I_{sp}$ depends on the ratio between thrust and mass flow. A heating case can reduce $\dot{m}$ and still increase $I_{sp}$ if the thrust per unit mass flow rises sufficiently. The data in Table~\ref{tab:performance} show exactly this behavior. The strongest heating case increases the total normalized thrust from 3.11 to 3.89 while the mass flux decreases only from $1.44$ to $1.39$ in units of $10^{-7}~{\rm kg\,s^{-1}}$. The mass-flow penalty is therefore modest compared with the impulse gain.

This does not mean that heating is universally beneficial. It means that the operating objective matters. If the desired quantity is total mass delivery, heating is unfavorable because it increases blockage. If the desired quantity is impulse per unit mass, strong heat addition can be favorable. The trade-off map in Fig.~\ref{fig:tradeoff} makes this distinction explicit. It also provides a more useful design metric than any single contour plot: the same thermal forcing that degrades $C_d$ can improve $I_{sp}$.

\subsection{Role of compression-zone restructuring}

The signed schlieren fields indicate that the compression signature also participates in the trade-off. Cooling maintains a sharper compression pattern, while heating distributes the compression over a broader region. In a continuum nozzle, a sharper shock would generally be associated with a strong irreversible loss. In the present rarefied nozzle, the compression region is already finite and wall affected. Heating does not simply move a shock upstream or downstream; it changes the relative importance of the core compression pattern and the wall thermal layer.

This is why the paper avoids using the phrase ``shock wave'' without qualification. The observable feature is a shock-cell-like compression structure. It has a finite width, it interacts with the diverging wall, and it changes together with the mass-flux thickness. The diagnostic statement is therefore:
\begin{equation}
\text{heating}\quad\Rightarrow\quad
\text{weaker signed density-gradient ridge}+\text{larger viscous--thermal 
layer}.
\end{equation}
This statement is consistent with both the schlieren visualization and the blockage profiles. It is also consistent with the $Kn_{\rm GLL}$ maps, which show that short-gradient-length non-equilibrium becomes more distributed under heating.

\subsection{Comparison with shock-centered low-rank nozzle data}

It is useful to compare the present POD result with the earlier shock-centered low-rank nozzle study. In that work, the walls were chosen to be specular in order to isolate the internal compression-layer dynamics from strong diffuse-wall accommodation. The dominant parametric variation was the location and finite thickness of an approximately normal internal compression layer. Once the layer was registered using the density-gradient station and a jump-based thickness, the density field became extremely compact: the leading centerline POD energy increased from 83.33\% in physical coordinates to 98.33\% in the jump-scaled coordinate, and the registered two-dimensional shock-window representation retained more than 99\% of its fluctuation energy in the first two modes.

The present case is different in two essential ways. First, the wall is thermally active, so the boundary condition changes the molecular energy distribution at the surface. Second, the wall-driven viscous--thermal layer occupies a substantial portion of the diverging passage. The flow response is therefore not only a translation and thickening of a compression layer. It also includes a change in the effective mass-flux passage and wall-adjacent non-equilibrium. It is therefore expected that a single registered mode is insufficient. The finding that four modes reconstruct representative schlieren fields with sub-percent error is still a strong compactness result, but it has a different meaning from the specular-wall case. It means that the thermally forced nozzle response is low-dimensional but multi-mechanism: compression-layer motion, wall-layer growth, outlet-cell reorganization and near-throat adjustment are all coherent but distinct components.

This comparison is important for reduced-order modelling. If one used the earlier specular-wall result alone, one might conclude that shock-centered registration is enough to collapse internal nozzle compression layers. The present heat-flux case shows the limitation of that conclusion. Registration of a compression layer is helpful, but active wall heat transfer introduces additional coherent modes. A robust surrogate or reduced-order representation for thermally forced rarefied nozzles should therefore include not only shock-position or shock-thickness coordinates, but also wall-thermal or blockage-related features. In this sense, the effective blockage metric is not only a performance diagnostic; it is also a candidate reduced coordinate for future operator-learning models.

\subsection{Interpretation of POD coefficients and finite-snapshot limitations}

The POD coefficient plot is intentionally not used as a monotonic law. The heat-flux sweep contains six snapshots, so the number of non-zero POD modes is necessarily limited. The POD energies should therefore be interpreted as compactness indicators for the chosen family of fields, not as a complete statistical basis. This is the same caution used in recent rarefied bow-shock POD analysis: low rank can partly reflect the size and construction of the parameter sweep, so the physical interpretation should focus on the organized leading patterns and on reconstruction quality, not on overinterpreting every high-order mode.

The coefficient trends also need not be monotonic. Each coefficient measures the projection of a field onto an orthogonal mode. When two physical effects change simultaneously--for example, wall-layer growth and outlet compression weakening--the projection on a given mode can pass through a local maximum or minimum. This is not a problem for the present argument. The important result is that the first few modes reconstruct the family accurately and that the spatial support of the modes coincides with the regions identified by schlieren, $Kn_{\rm GLL}$ and blockage diagnostics. POD therefore acts as an independent structural check: the heat-flux response is not random DSMC scatter and not a collection of unrelated fields, but a coherent deformation of the compression and wall-layer structure.

\subsection{Relation to heat-transfer and propulsion design}

From a heat-transfer perspective, the most important outcome is that the wall heat flux does not remain localized as a wall-temperature response. It propagates into the flow through the mass-flux distribution, the density-gradient field and the propulsion metrics. The wall heat flux, therefore, couples the thermal and aerodynamic design problems. An engineering design based only on fixed wall temperature would miss this coupling, because the wall temperature distribution is itself the result of the prescribed energy balance. Conversely, a design based only on global thrust or mass flow would miss the internal reason for the performance trend: the contraction of the effective mass-flux thickness.

For micro-propulsion applications, the result suggests two possible operating strategies. In a mass-flow-limited system, cooling or weak heating may be preferred because it preserves a wider effective passage. In an impulse-limited system, strong heating may be attractive because it increases thrust per unit mass flow. The present simulations do not include material constraints, conjugate heat transfer or wall-temperature limits, so they do not define an optimum device. They do, however, provide the physical structure of the trade-off that such an optimization would have to consider.

\subsection{Limitations and robustness of the interpretation}

Several limitations should be stated explicitly. First, the present geometry is planar and two-dimensional. The blockage metric is therefore defined per unit depth and should not be transferred directly to an axisymmetric nozzle without modifying the area weighting. In an axisymmetric geometry, the mass-flux thickness would need to include the radial Jacobian and the wall curvature would modify the near-wall thermal layer. Second, the heat-flux boundary condition is imposed on the gas side without solving a conjugate wall-conduction problem. The wall-temperature distribution obtained here is therefore the gas-dynamic wall temperature required to realize the target flux, not the temperature of a finite-conductivity solid wall. Third, the gas is nitrogen without vibrational excitation or chemistry. At much higher temperatures or in chemically active micro-thrusters, vibrational and chemical energy modes could modify the enthalpy partition and the resulting impulse response.

These limitations do not weaken the main conclusion because the central mechanism is based on directly resolved DSMC quantities: $\rho u$, wall-temperature response, signed density gradient, and integrated thrust. The effective blockage metric does not depend on a continuum boundary-layer assumption, and the trade-off is observed simultaneously in the field-level and integral-level diagnostics. The result should therefore be understood as a controlled kinetic demonstration of a thermal--aerodynamic mechanism. Future work should test how the numerical values of $\beta_m$, $C_d$, and $I_{sp}$ change when gas-surface accommodation, wall conduction, gas species, and three-dimensional geometry are varied. The expected qualitative mechanism remains clear: any thermal boundary condition that thickens a low-momentum wall layer will reduce the effective mass-carrying core, while any sufficiently strong heat addition can increase impulse if it raises thrust more than it reduces mass flow.

The interpretation of POD is similarly bounded. POD is not used here as evidence of turbulence coherence or as a predictive reduced-order model. It is a parameter-space compactness diagnostic of six steady DSMC fields. The finite number of snapshots limits the maximum number of non-zero modes, and the mode signs are arbitrary. The useful information is therefore the rapid decay of cumulative energy, the spatial localization of the leading modes, and the successful reconstruction of representative cases with a common four-mode basis. These three facts support the statement that heat flux reorganizes the schlieren field within a small number of coherent patterns. They do not imply that every future heat-flux value or every geometry will be represented by the same four modes without retraining or resampling.

\section{Conclusions}
\label{sec:conclusions}

A DSMC study was performed to examine prescribed wall-heat-flux effects in a nitrogen converging--diverging micro-nozzle. The heat flux was applied to the diverging wall and included cooling, adiabatic, and heating cases. The analysis combines wall and bulk temperature fields with heat-transfer scaling, rarefaction diagnostics, mass-flux-thickness analysis, propulsion metrics, signed numerical schlieren, and POD. The main conclusions are as follows.

\begin{enumerate}
\item Prescribed wall heat flux produces a coupled but non-parallel wall--bulk thermal response. Strong heating raises $T_w/T_0$ above five near the exit, whereas the mass-flux-weighted bulk temperature first decreases through expansion and only recovers downstream for sufficiently large heat input. The resulting wall--bulk stratification is the thermal origin of the blockage and impulse trends.

\item The first-cell near-wall tangential slip response confirms that wall heating changes the momentum exchange at the gas--surface interface. Strong heating maintains positive $u_{\rm slip}/U_{in,ad}$ over a longer portion of Wall-2, supporting the interpretation that a wall-driven viscous--thermal layer contracts the effective mass-carrying passage.

\item The apparent Nusselt and local-viscosity Brinkman diagnostics must be interpreted as heat-flux-based scalings rather than continuum Fourier-law measurements. The Nusselt-type response is evaluated with a film-temperature VHS/Wu--Eucken conductivity, $k_f=f_{{\rm eu},N_2}(f_{\rm dof}/2)(k_B/m)\mu_f$, using $f_{{\rm eu},N_2}=1.96$ and $f_{\rm dof}=5$ for non-vibrating nitrogen. Cooling cases contain locations where $T_w-T_b=0$, so $Nu_q^{\rm loc}$ becomes singular; the raw singular behavior is retained for diagnosis and a validity mask is used only for finite-ratio comparisons. Heating cases have valid, low-amplitude $Nu_q^{\rm loc}$ values because the wall--bulk temperature difference and $k_f$ both grow with heat input. The Brinkman-type ratio is evaluated with the VHS temperature-dependent viscosity, $\mu_b=\mu_{\rm ref}(T_b/T_{\rm ref})^\omega$, and decreases with heating even after the local viscosity increase is included; this confirms the transition from momentum-dominated response to thermal augmentation.

\item The internal compression feature is best interpreted as a finite shock-cell-like viscous--thermal compression zone rather than a continuum discontinuity. Signed numerical schlieren shows that heating weakens and spreads this structure.

\item The mass-flux-thickness metric $\beta_m=h_m/h_{\rm geo}$ provides a direct DSMC-field measure of the effective mass-carrying core. Strong heating reduces $\beta_m$ and increases the effective blockage $1-\beta_m$ in the diverging section.

\item The integrated response is a thermal--aerodynamic trade-off. Wall heating reduces discharge coefficient and mass flow rate, but strong heating increases specific impulse. The strongest heating case reaches $I_{sp}=201~{\rm s}$ compared with $156~{\rm s}$ in the adiabatic case.

\item Gradient-length Knudsen-number diagnostics show that heating redistributes local non-equilibrium from a compact compression signature toward a broader wall-driven viscous--thermal layer.

\item POD of the signed numerical-schlieren field shows that the heat-flux family is compact but not trivially one-dimensional. The first two modes capture more than $97\%$ of the fluctuation energy, and four modes reconstruct representative cases with sub-percent error. Compared with shock-centered specular-wall nozzle data, the need for additional modes reflects the extra coherent degree of freedom introduced by wall thermal forcing. The POD result should be interpreted as compactness of the present six-member heat-flux family, not as a universal reduced basis for all thermally forced nozzle configurations.
\end{enumerate}

The principal physical message is that prescribed wall heat flux modifies rarefied micro-nozzle operation by changing the aerodynamic passage available to the gas. The wall does not act only as a source or sink of thermal energy; it changes the effective mass-flux thickness, local non-equilibrium structure, wall-to-bulk thermal coupling, compression-layer organization, and propulsion efficiency. Future work should examine whether the same trade-off persists under different accommodation coefficients, gas species, inlet Knudsen numbers, three-dimensional geometries, and conjugate wall-thermal models.

\appendix
\section{Raw signed local Nusselt-type response}
\label{app:raw_nusselt}

\begin{figure}[!htbp]
\centering
\includegraphics[width=0.98\textwidth]{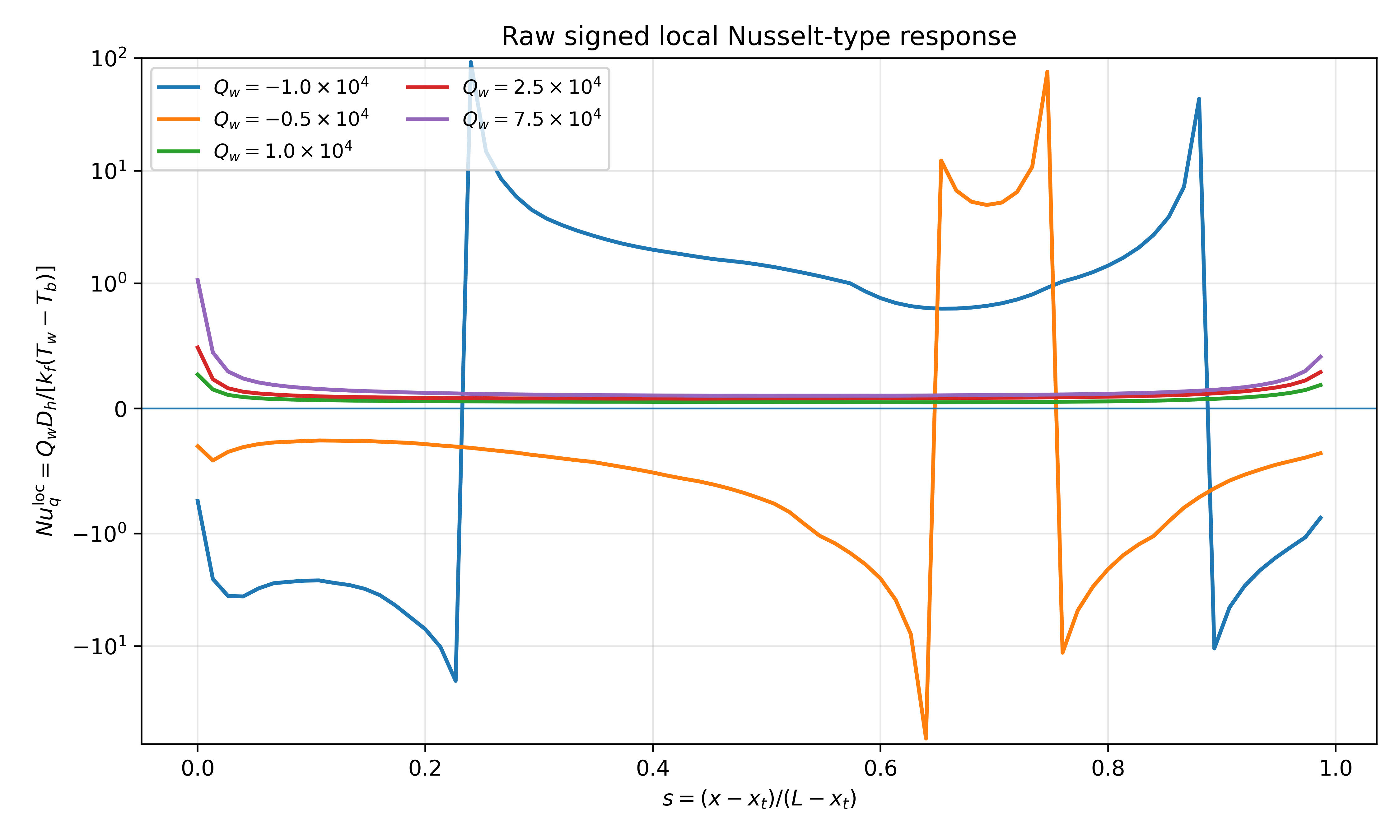}
\caption{Raw signed local Nusselt-type response on Wall-2 using the film-temperature VHS/Eucken conductivity. The signed-symmetric logarithmic scale retains the singular excursions associated with wall--bulk temperature crossings. These excursions motivate the validity mask used in Fig.~\ref{fig:nuvalid}.}
\label{fig:raw_nusselt}
\end{figure}

Figure~\ref{fig:raw_nusselt} shows the raw signed $Nu_q^{\rm loc}$ profile before applying the singular-denominator validity mask. The sharp excursions occur where $T_w-T_b$ approaches zero, especially in the cooling cases. They are therefore an expected consequence of the imposed-flux-to-temperature-difference definition, not a numerical smoothing failure. The corresponding finite-ratio comparison in the main text masks only the narrow band $|T_w-T_b|\le0.05T_0$.

\section*{Data availability}

The DSMC field data and post-processing scripts used to compute the blockage, near-wall slip, numerical-schlieren, $Kn_{\rm GLL}$ and POD diagnostics can be made available by the corresponding author upon reasonable request.

\section*{Declaration of competing interest}

The authors declare no competing interests.

\clearpage
\bibliographystyle{elsarticle-harv}
\bibliography{references}

@article{tsien1946,
  author  = {Tsien, Hsue-Shen},
  title   = {Superaerodynamics, Mechanics of Rarefied Gases},
  journal = {Journal of the Aeronautical Sciences},
  year    = {1946},
  volume  = {13},
  number  = {12},
  pages   = {653--664},
  doi     = {10.2514/8.11476}
}

@book{schaaf1961,
  author    = {Schaaf, Samuel A. and Chambr\'e, Paul L.},
  title     = {Flow of Rarefied Gases},
  publisher = {Princeton University Press},
  year      = {1961},
  doi       = {10.1515/9781400885800}
}

@book{karniadakis2005,
  author    = {Karniadakis, George E. and Beskok, Ali and Aluru, Narayan R.},
  title     = {Microflows and Nanoflows: Fundamentals and Simulation},
  publisher = {Springer},
  address   = {New York},
  year      = {2005},
  doi       = {10.1007/0-387-28676-4}
}

@article{bird1970,
  author  = {Bird, G. A.},
  title   = {Direct Simulation and the Boltzmann Equation},
  journal = {The Physics of Fluids},
  year    = {1970},
  volume  = {13},
  number  = {11},
  pages   = {2676--2681},
  doi     = {10.1063/1.1692849}
}

@article{wagner1992,
  author  = {Wagner, Wolfgang},
  title   = {A convergence proof for Bird's direct simulation Monte Carlo method for the Boltzmann equation},
  journal = {Journal of Statistical Physics},
  year    = {1992},
  volume  = {66},
  number  = {3--4},
  pages   = {1011--1044},
  doi     = {10.1007/BF01055714}
}

@inproceedings{ivanov1999,
  author    = {Ivanov, M. S. and Markelov, G. N. and Ketsdever, A. D. and Wadsworth, D. C.},
  title     = {Numerical Study of Cold Gas Micronozzle Flows},
  booktitle = {37th Aerospace Sciences Meeting and Exhibit},
  year      = {1999},
  pages     = {1--11},
  doi       = {10.2514/6.1999-166}
}

@article{wang2004,
  author  = {Wang, Moran R. and Li, Z. X.},
  title   = {Numerical simulations on performance of MEMS-based nozzles at moderate or low temperatures},
  journal = {Microfluidics and Nanofluidics},
  year    = {2004},
  volume  = {1},
  number  = {1},
  pages   = {62--70},
  doi     = {10.1007/s10404-004-0008-5}
}

@article{hao2005,
  author  = {Hao, Peng-Fei and Ding, Yong-Tang and Yao, Zhao-Hui and He, Feng and Zhu, Ke-Qin},
  title   = {Size effect on gas flow in micro nozzles},
  journal = {Journal of Micromechanics and Microengineering},
  year    = {2005},
  volume  = {15},
  number  = {11},
  pages   = {2069--2073},
  doi     = {10.1088/0960-1317/15/11/019}
}

@article{xie2007,
  author  = {Xie, Chong},
  title   = {Characteristics of micronozzle gas flows},
  journal = {Physics of Fluids},
  year    = {2007},
  volume  = {19},
  number  = {3},
  pages   = {037102},
  doi     = {10.1063/1.2709707}
}

@article{horisawa2008,
  author  = {Horisawa, Hideyuki and Sawada, Fujimi and Onodera, Kosuke and Funaki, Ikkoh},
  title   = {Numerical simulation of micro-nozzle and micro-nozzle-array flowfield characteristics},
  journal = {Vacuum},
  year    = {2008},
  volume  = {83},
  number  = {1},
  pages   = {52--56},
  doi     = {10.1016/j.vacuum.2008.03.097}
}

@article{darbandi2011,
  author  = {Darbandi, Masoud and Roohi, Ehsan},
  title   = {Study of subsonic--supersonic gas flow through micro/nanoscale nozzles using unstructured DSMC solver},
  journal = {Microfluidics and Nanofluidics},
  year    = {2011},
  volume  = {10},
  number  = {2},
  pages   = {321--335},
  doi     = {10.1007/s10404-010-0671-7}
}

@article{lijo2015,
  author  = {Lijo, V. and Setoguchi, T. and Kim, Heuy Dong},
  title   = {Analysis of Supersonic Micronozzle Flows},
  journal = {Journal of Propulsion and Power},
  year    = {2015},
  volume  = {31},
  number  = {2},
  pages   = {754--757},
  doi     = {10.2514/1.B35453}
}

@article{saadati2015,
  author  = {Saadati, Seyed Ali and Roohi, Ehsan},
  title   = {Detailed investigation of flow and thermal field in micro/nano nozzles using Simplified Bernoulli Trial (SBT) collision scheme in DSMC},
  journal = {Aerospace Science and Technology},
  year    = {2015},
  volume  = {46},
  pages   = {236--255},
  doi     = {10.1016/j.ast.2015.07.013}
}

@article{sabouri2019,
  author  = {Sabouri, Moslem and Darbandi, Masoud},
  title   = {Numerical study of species separation in rarefied gas mixture flow through micronozzles using DSMC},
  journal = {Physics of Fluids},
  year    = {2019},
  volume  = {31},
  number  = {4},
  pages   = {042004},
  doi     = {10.1063/1.5083807}
}

@article{mahdavi2020,
  author  = {Mahdavi, Amirmehran and Roohi, Ehsan},
  title   = {A novel hybrid DSMC-Fokker Planck algorithm implemented to rarefied gas flows},
  journal = {Vacuum},
  year    = {2020},
  volume  = {181},
  pages   = {109736},
  doi     = {10.1016/j.vacuum.2020.109736}
}

@article{sukesan2021,
  author  = {Sukesan, Manu K. and Shine, S. R.},
  title   = {Geometry effects on flow characteristics of micro-scale planar nozzles},
  journal = {Journal of Micromechanics and Microengineering},
  year    = {2021},
  volume  = {31},
  number  = {12},
  pages   = {125001},
  doi     = {10.1088/1361-6439/ac2bac}
}

@article{kosyanchuk2021,
  author  = {Kosyanchuk, V. V. and Yakunchikov, A. N.},
  title   = {Separation of a binary gas mixture outflowing into vacuum through a micronozzle},
  journal = {Physics of Fluids},
  year    = {2021},
  volume  = {33},
  number  = {8},
  pages   = {082007},
  doi     = {10.1063/5.0055879}
}

@article{groll2023,
  author  = {Groll, Rodion and Frieler, Tobias},
  title   = {Validation of DSMC mass flow modeling for transsonic gas flows in micro-propulsion systems},
  journal = {Frontiers in Mechanical Engineering},
  year    = {2023},
  volume  = {9},
  pages   = {1217645},
  doi     = {10.3389/fmech.2023.1217645}
}

@article{zhang2024,
  author  = {Zhang, Shurui and Li, Yong and Wang, Xudong and Lu, Songcai and Yu, Yusong and Yang, Jun},
  title   = {Effects of the Wall Temperature on Rarefied Gas Flows and Heat Transfer in a Micro-Nozzle},
  journal = {Micromachines},
  year    = {2024},
  volume  = {15},
  number  = {1},
  pages   = {22},
  doi     = {10.3390/mi15010022}
}

@article{sabouri2024,
  author  = {Sabouri, Moslem and Zakeri, Ramin and Ebrahimi, Amin},
  title   = {Improving computational efficiency in DSMC simulations of vacuum gas dynamics with a fixed number of particles per cell},
  journal = {Physica Scripta},
  year    = {2024},
  volume  = {99},
  number  = {8},
  pages   = {085213},
  doi     = {10.1088/1402-4896/ad5a46}
}

@article{akhlaghi2012,
  author  = {Akhlaghi, Hassan and Roohi, Ehsan and Stefanov, Stefan},
  title   = {A new iterative wall heat flux specifying technique in DSMC for heating/cooling simulations of MEMS/NEMS},
  journal = {International Journal of Thermal Sciences},
  year    = {2012},
  volume  = {59},
  pages   = {111--125},
  doi     = {10.1016/j.ijthermalsci.2012.04.002}
}

@article{akhlaghi2016,
  author  = {Akhlaghi, Hassan and Roohi, Ehsan},
  title   = {A novel algorithm for implementing a specified wall heat flux in DSMC: Application to micro/nano flows and hypersonic flows},
  journal = {Computers \& Fluids},
  year    = {2016},
  volume  = {127},
  pages   = {78--101},
  doi     = {10.1016/j.compfluid.2015.12.008}
}

@article{louisos2012,
  author  = {Louisos, William F. and Hitt, Darren L.},
  title   = {Viscous Effects on Performance of Three-Dimensional Supersonic Micronozzles},
  journal = {Journal of Spacecraft and Rockets},
  year    = {2012},
  volume  = {49},
  number  = {1},
  pages   = {51--58},
  doi     = {10.2514/1.53026}
}

@article{rothe1971,
  author  = {Rothe, D. E.},
  title   = {Electron-Beam Studies of Viscous Flow in Supersonic Nozzles},
  journal = {AIAA Journal},
  year    = {1971},
  volume  = {9},
  number  = {5},
  pages   = {804--811},
  doi     = {10.2514/3.49904}
}

@article{peyvan2024,
  title={Fusion-deeponet: A data-efficient neural operator for geometry-dependent hypersonic and supersonic flows},
  author={Peyvan, Ahmad and Kumar, Varun and Karniadakis, George Em},
  journal={Journal of Computational Physics},
  pages={114432},
  year={2026},
  publisher={Elsevier}
}

@article{lockerby2004velocity,
  author  = {Lockerby, Duncan A. and Reese, Jason M. and Emerson, David R. and Barber, Robert W.},
  title   = {Velocity boundary condition at solid walls in rarefied gas calculations},
  journal = {Physical Review E},
  year    = {2004},
  volume  = {70},
  number  = {1},
  pages   = {017303},
  doi     = {10.1103/PhysRevE.70.017303}
}

@article{maxwell1879,
  author  = {Maxwell, James Clerk},
  title   = {On stresses in rarified gases arising from inequalities of temperature},
  journal = {Philosophical Transactions of the Royal Society of London},
  year    = {1879},
  volume  = {170},
  pages   = {231--256},
  doi     = {10.1098/rstl.1879.0067}
}

@article{raissi2019,
  author  = {Raissi, Maziar and Perdikaris, Paris and Karniadakis, George Em},
  title   = {Physics-informed neural networks: A deep learning framework for solving forward and inverse problems involving nonlinear partial differential equations},
  journal = {Journal of Computational Physics},
  year    = {2019},
  volume  = {378},
  pages   = {686--707},
  doi     = {10.1016/j.jcp.2018.10.045}
}

@article{lu2021deeponet,
  author  = {Lu, Lu and Jin, Pengzhan and Pang, Guofei and Zhang, Zhongqiang and Karniadakis, George Em},
  title   = {Learning nonlinear operators via DeepONet based on the universal approximation theorem of operators},
  journal = {Nature Machine Intelligence},
  year    = {2021},
  volume  = {3},
  number  = {3},
  pages   = {218--229},
  doi     = {10.1038/s42256-021-00302-5}
}

@article{tatsios2025,
  author  = {Tatsios, Georgios and others},
  title   = {A DSMC-CFD coupling method using surrogate modelling for low-speed rarefied gas flows},
  journal = {Journal of Computational Physics},
  year    = {2025},
  volume  = {520},
  pages   = {113500},
  doi     = {10.1016/j.jcp.2024.113500}
}

@article{roohi2026microstepdeeponet,
  author  = {Roohi, Ehsan and Mahdavi, Amirmehran},
  title   = {Analysis of the rarefied flow at micro-step using a DeepONet surrogate model with a physics-guided zonal loss function},
  journal = {Microfluidics and Nanofluidics},
  year    = {2026},
  volume  = {30},
  number  = {44},
  pages   = {44},
  doi     = {10.1007/s10404-026-02899-8}
}

@article{roohi2025shockfusion,
  title={Shock-Centered Low-Rank Structure and Neural-Operator Representation of Rarefied Micro-Nozzle Flows},
  author={Roohi, Ehsan and Mahdavi, Amirmehran},
  journal={arXiv preprint arXiv:2605.12723},
  year={2026}
}

@article{agir2022rarefaction_edney,
  author  = {Agir, Muhammed Burak and White, Craig and Kontis, Konstantinos},
  title   = {The effect of increasing rarefaction on the formation of Edney shock interactions},
  journal = {Shock Waves},
  year    = {2022},
  doi     = {10.1007/s00193-022-01109-y}
}

@inproceedings{riabov1999shock_interference_rarefied,
  author    = {Riabov, Vladimir V.},
  title     = {Shock Interference in Hypersonic Rarefied-Gas Flows Near a Cylinder},
  booktitle = {AIAA 30th Fluid Dynamics Conference},
  year      = {1999},
  doi       = {10.2514/6.1999-3207}
}

@article{sharipov2011,
  author  = {Sharipov, Felix},
  title   = {Data on the Velocity Slip and Temperature Jump on a Gas-Solid Interface},
  journal = {Journal of Physical and Chemical Reference Data},
  year    = {2011},
  volume  = {40},
  number  = {2},
  pages   = {023101},
  doi     = {10.1063/1.3580290}
}

@article{balaj2014cwh,
  author  = {Balaj, Mojtaba and Roohi, Ehsan and Akhlaghi, Hassan and Myong, Rho Shin},
  title   = {Investigation of convective heat transfer through constant wall heat flux micro/nano channels using DSMC},
  journal = {International Journal of Heat and Mass Transfer},
  year    = {2014},
  volume  = {71},
  pages   = {633--638},
  doi     = {10.1016/j.ijheatmasstransfer.2013.12.053}
}

@article{balaj2015shearwork,
  author  = {Balaj, Mojtaba and Roohi, Ehsan and Akhlaghi, Hassan},
  title   = {Effects of shear work on non-equilibrium heat transfer characteristics of rarefied gas flows through micro/nanochannels},
  journal = {International Journal of Heat and Mass Transfer},
  year    = {2015},
  volume  = {83},
  pages   = {69--74},
  doi     = {10.1016/j.ijheatmasstransfer.2014.11.087}
}

@article{wu2015polyatomic,
  author  = {Wu, Lei and White, Craig and Scanlon, Thomas J. and Reese, Jason M. and Zhang, Yonghao},
  title   = {A kinetic model of the Boltzmann equation for non-vibrating polyatomic gases},
  journal = {Journal of Fluid Mechanics},
  year    = {2015},
  volume  = {763},
  pages   = {24--50},
  doi     = {10.1017/jfm.2014.632}
}

@article{wu2020eucken,
  author  = {Wu, Lei and Li, Qi and Liu, Haihu and Ubachs, Wim},
  title   = {Extraction of the translational Eucken factor from light scattering by molecular gas},
  journal = {Journal of Fluid Mechanics},
  year    = {2020},
  volume  = {901},
  pages   = {A23},
  doi     = {10.1017/jfm.2020.568}
}

\end{document}